\pdfoutput=1
\documentclass[aps,prd,superscriptaddress,showkeys,twocolumn,nofootinbib,10pt,floatfix]{revtex4-2}
\usepackage{preamble}

\begin{document}

\title{Ultralight scalar field domain walls and Galileo GNSS atomic clocks}

\author{Alessandro Di Marco\orcidA{}}
\email{alessandro.dimarco1@inaf.it}
\affiliation{Istituto Nazionale di Astrofisica, Istituto di Astrofisica e Planetologia Spaziali (INAF-IAPS), Via Fosso del Cavaliere, 100, 00133 Roma, Italy}
\affiliation{Istituto Nazionale di Fisica Nucleare (INFN), Sezione di Tor Vergata, Via della Ricerca Scientifica 1, 00133 Roma, Italy}

\author{Feliciana Sapio\orcidB{}}
\affiliation{Istituto Nazionale di Astrofisica, Istituto di Astrofisica e Planetologia Spaziali (INAF-IAPS), Via Fosso del Cavaliere, 100, 00133 Roma, Italy}
\affiliation{Istituto Nazionale di Fisica Nucleare (INFN), Sezione di Tor Vergata, Via della Ricerca Scientifica 1, 00133 Roma, Italy}

\author{Massimo Visco\orcidC{}}
\affiliation{Istituto Nazionale di Astrofisica, Istituto di Astrofisica e Planetologia Spaziali (INAF-IAPS), Via Fosso del Cavaliere, 100, 00133 Roma, Italy}
\affiliation{Istituto Nazionale di Fisica Nucleare (INFN), Sezione di Tor Vergata, Via della Ricerca Scientifica 1, 00133 Roma, Italy}

\author{David Lucchesi\orcidD{}}
\affiliation{Istituto Nazionale di Astrofisica, Istituto di Astrofisica e Planetologia Spaziali (INAF-IAPS), Via Fosso del Cavaliere, 100, 00133 Roma, Italy}
\affiliation{Istituto Nazionale di Fisica Nucleare (INFN), Sezione di Tor Vergata, Via della Ricerca Scientifica 1, 00133 Roma, Italy}

\author{Marco Cinelli\orcidE{}}
\affiliation{Istituto Nazionale di Astrofisica, Istituto di Astrofisica e Planetologia Spaziali (INAF-IAPS), Via Fosso del Cavaliere, 100, 00133 Roma, Italy}
\affiliation{Istituto Nazionale di Fisica Nucleare (INFN), Sezione di Tor Vergata, Via della Ricerca Scientifica 1, 00133 Roma, Italy}

\author{Emiliano Fiorenza\orcidF{}}
\affiliation{Istituto Nazionale di Astrofisica, Istituto di Astrofisica e Planetologia Spaziali (INAF-IAPS), Via Fosso del Cavaliere, 100, 00133 Roma, Italy}

\author{Carlo Lefevre\orcidG{}}
\affiliation{Istituto Nazionale di Astrofisica, Istituto di Astrofisica e Planetologia Spaziali (INAF-IAPS), Via Fosso del Cavaliere, 100, 00133 Roma, Italy}

\author{Pasqualino Loffredo\orcidH{}}
\affiliation{Istituto Nazionale di Astrofisica, Istituto di Astrofisica e Planetologia Spaziali (INAF-IAPS), Via Fosso del Cavaliere, 100, 00133 Roma, Italy}

\author{Marco Lucente\orcidI{}}
\affiliation{Istituto Nazionale di Astrofisica, Istituto di Astrofisica e Planetologia Spaziali (INAF-IAPS), Via Fosso del Cavaliere, 100, 00133 Roma, Italy}
\affiliation{Istituto Nazionale di Fisica Nucleare (INFN), Sezione di Tor Vergata, Via della Ricerca Scientifica 1, 00133 Roma, Italy}

\author{Carmelo Magnafico\orcidL{}}
\affiliation{Istituto Nazionale di Astrofisica, Istituto di Astrofisica e Planetologia Spaziali (INAF-IAPS), Via Fosso del Cavaliere, 100, 00133 Roma, Italy}
\affiliation{Istituto Nazionale di Fisica Nucleare (INFN), Sezione di Tor Vergata, Via della Ricerca Scientifica 1, 00133 Roma, Italy}

\author{Roberto Peron\orcidM{}}
\affiliation{Istituto Nazionale di Astrofisica, Istituto di Astrofisica e Planetologia Spaziali (INAF-IAPS), Via Fosso del Cavaliere, 100, 00133 Roma, Italy}
\affiliation{Istituto Nazionale di Fisica Nucleare (INFN), Sezione di Tor Vergata, Via della Ricerca Scientifica 1, 00133 Roma, Italy}

\author{Francesco Santoli\orcidN{}}
\affiliation{Istituto Nazionale di Astrofisica, Istituto di Astrofisica e Planetologia Spaziali (INAF-IAPS), Via Fosso del Cavaliere, 100, 00133 Roma, Italy}

\author{Francesco Vespe\orcidO{}}
\affiliation{Centro di Geodesia Spaziale (CGS) ``G. Colombo'', Agenzia Spaziale Italiana (ASI), 75100 Matera, Italy}

\date{\today}

\begin{abstract}

In this article, we study the impact of an Earth-sized ultralight scalar field domain wall (DW) 
on the European Galileo Global Navigation Satellite System constellation 
and the feasibility of a related detection. 
In particular, a set of statistical procedures is developed and implemented 
to characterize the statistical properties of the background noise for a subnetwork of $N$ Galileo satellite clocks, 
including a false-alarm rate estimation and a detection efficiency analysis. 
These methods allow to implement a coincidence analysis and establish a preliminary statistical framework for a robust and controlled search 
for domain-wall crossings in clock data, and for deriving constraints 
on the associated fundamental particle physics parameters. 
The activity is addressed from the point of
view of the scientific objectives of the Galileo for Science (G4S$\_2.0$) project.

\end{abstract}

\maketitle
\tableofcontents

\section{Introduction}
\label{introduction}

The behavior of the extremely high-energy universe remains largely unknown and is the subject of intense theoretical speculation. 
However, there is broad consensus within the particle physics and cosmological communities that, 
at energy scales well below the Planck scale, the universe is expected to have evolved through the unbroken phase of the Standard Model (SM) of particle physics,
described by the gauge symmetry group $SU(3)_C \times SU(2)_L \times U(1)_Y$.
As the universe cooled down to the critical temperature of the model, 
the Higgs-driven phase transition took place and triggered the spontaneous breaking of the electroweak symmetry $SU(2)_L \times U(1)_Y$ down to the electromagnetic one $U(1)_{\mathrm{EM}}$. 
As a consequence, the universe entered the broken phase of the SM, characterized by the residual symmetry $SU(3)_C \times U(1)_{\mathrm{EM}}$, which defines the present vacuum state of the universe 
(see \cite{Burgess:2006hbd,Thomson:2013zua,Kane:2017aaa} for introductory overview on the SM
and \cite{Donoghue:1992dd} for more technical details).
However, the thermal history of the universe may also have included one or more phase transitions, typically involving
additional scalar sectors that are weakly coupled or entirely decoupled from the SM. 
Such transitions can be associated with non-trivial symmetry-breaking patterns and may lead to relevant cosmological implications, including the formation of cosmic defects.

In general, a phase transition characterized by the spontaneous breaking of a symmetry group $G$ to a subgroup $H$ leads to a vacuum manifold $\mathcal{V} = G/H$. 
If this manifold has a non-trivial homotopy group, relic topological defects generically form. 
In particular, if the zeroth homotopy group of the vacuum manifold is non-trivial, $\pi_0(\mathcal{V}) \neq 0$, 
\textit{namely} if the vacuum consists of disconnected minima, the formation of a network of domain walls (DWs) is expected.
The emergence of cosmological domain structures was first explored in the context of spontaneous $CP$ symmetry breaking in the early universe by 
\emph{Lee} and \emph{Zeldovich}
\cite{Lee:1974jb,Zeldovich:1974uw}. 
Subsequently, \cite{Kibble:1976sj,Kibble:1980mv,Lazarides:1981fv} 
provided the first comprehensive analyses of the domain wall properties and formation mechanisms, together with other topological defects.
For complete reviews, see \cite{Vilenkin:2000jqa,Vachaspati:2006zz}.
The most standard example is the spontaneous breaking of a discrete symmetry $G = \mathbb{Z}_N$ to \emph{nothing}. 
The minimal realization of this scenario corresponds to the reflection symmetry $G = \mathbb{Z}_2$, for which the vacuum manifold $\mathcal{V}$ consists of a pair of distinct, degenerate minima.

In the case of first-order phase transitions, domain walls typically form through bubble nucleation and subsequent collisions, 
while in continuous or second-order phase transitions their formation proceeds via the 
original \emph{Kibble--Zurek mechanism} \cite{Vachaspati:2006zz}.

The subsequent evolution of a domain wall network is highly non-trivial and exhibits a rich dynamical structure.
Typically, shortly after formation, the number of domain walls per Hubble volume is very large, 
and their dynamics is strongly influenced by interactions with the surrounding relativistic plasma.
In this early stage, frictional forces dominate and significantly damp the motion of the walls.

As the universe expands and the temperature of the thermal plasma decreases, friction becomes progressively negligible.
The DW network is then expected to approach an attractor solution, a \emph{linear scaling regime}, 
in which the number of domain walls per Hubble volume becomes $\sim\mathcal{O}(1)$ and the average wall velocity approaches relativistic values. 
In this regime, the energy density of the domain wall network scales with cosmic time $t_C$ as $\propto t_C^{-1}$,
decreasing more slowly than the background energy density associated with radiation or matter \cite{Zeldovich:1974uw}.

As a consequence, the domain wall network inevitably comes to dominate the total energy density of the universe in a finite time. 
The corresponding equation of state of the domain wall distribution leads to an accelerated cosmic expansion, giving rise to a catastrophic cosmological scenario usually denoted as the \emph{domain wall catastrophe} \cite{Zeldovich:1974uw}.

It is also important to stress that, under certain conditions, the domain wall network may eventually dominate the cosmic energy density even without ever becoming relativistic and/or entering a linear scaling regime.

However, the occurrence of a domain-wall-dominated phase is incompatible with current cosmological and astrophysical observations, 
indicating that the standard picture must be modified and that domain walls, once formed, 
should eventually disappear in a finite time, possibly before the current cosmological epoch \cite{Zeldovich:1974uw}.
Alternatively, the domain wall distribution should represent an almost negligible contribution to the global energy density.
The literature contains a large number of theoretical proposals 
addressing the domain wall problem.
In general, the gravitational properties of (planar) domain walls
and, in particular, the possibility of their instability and 
subsequent collapse were primarily studied by \emph{Vilenkin} \cite{Vilenkin:1981zs,Vilenkin:1983ykn}.
However, simple scalar field domain walls could disappear due to 
cosmic string-boundaries and holes \cite{Kibble:1982dd,Vilenkin:1982ks},
low-temperature symmetry restoration \cite{Kuzmin:1981ip},
dynamically induced temperature-dependent bias in the scalar potential \cite{Mohanty:1984pj,Tamvakis:1981aaa},
or the spontaneous breakdown of anomalous discrete symmetries \cite{Preskill:1991kd}, just to name a few.
However, the most widely studied resolution simply consists in assuming that the underlying domain wall particle physics model
is based on a discrete symmetry which is \emph{explicitly} and \emph{slightly} broken.
In this sense, a small \textit{a priori} explicit symmetry breaking term in the scalar potential lifts the degeneracy of the vacua, 
selects a preferred minimum, and induces a pressure difference across each wall. 
As a result, domain walls collapse and effectively decay before they can dominate the energy density of the universe.
Note that QCD axion or axion-like particle cosmology
can also experience a domain wall problem 
-- although with crucial wall-production differences --
whose solution may rely on similar mechanisms; see, for instance, \cite{Vilenkin:1982ks,Sikivie:1982qv}.

The discussed \emph{bias} solution is supported by the properties of gravity \cite{Rai:1992xw} 
and, in particular, by the fact that
a hypothetical quantum gravity theory, such as (super) string theory, naturally breaks exact global symmetries 
\cite{Giddings:1987cg,Banks:1988yz,Kallosh:1995hi,Banks:2010zn},
as also emerged in the context of the swampland program \cite{Vafa:2005ui,Arkani-Hamed:2006emk,Ooguri:2006in,Fichet:2019ugl,Harlow:2018tng,Daus:2020vtf}.
The detailed cosmological evolution of a domain wall network was first investigated by \emph{Press et al.} in \cite{Press:1989yh},
while biased scenarios were studied in \cite{Gelmini:1988sf,Coulson:1995nv,Larsson:1996sp}.
More recently, joint efforts between 
analytical \cite{Hindmarsh:1996xv,Hindmarsh:2002bq,Avelino:2005kn,Martins:2016lzc,Avelino:2019wqd,Avelino:2020ubr,Avelino:2022zem}
and numerical \cite{Garagounis:2002kt,Oliveira:2004he,Avelino:2005pe,Leite:2011sc,Leite:2012vn,Correia:2014kqa,Martins:2016ois,Correia:2018tty}
investigations have suggested that the evolution of a domain wall network can be satisfactorily described
by the so-called \emph{velocity-dependent one-scale model} (VOS) or its direct extensions.

In cosmological scenarios featuring non-catastrophic domain walls (or axion domain walls), 
a number of interesting phenomenological consequences may arise. 
In such a regime, DWs could provide an explanation for the hypothetical dark sector of the universe
\cite{Bucher:1998mh,Battye:1999eq,Friedland:2002qs,PinaAvelino:2006ia,Avelino:2006xy,Avelino:2008ve,Avelino:2008qy},
predict specific anisotropies in the CMB power spectrum 
\cite{Sousa:2015cqa,Lazanu:2015fua},
source cosmic birefringence in the CMB
\cite{Takahashi:2020tqv,Fujita:2020ecn,Gonzalez:2022mcx,Ferreira:2023jbu,Lee:2025yvn},
trigger the observed matter-antimatter or baryon asymmetry of the universe
\cite{Brandenberger:1994mq,Abel:1995uc,Prokopec:1995gp,Brandenberger:1998kn,Daido:2015gqa,Daido:2015bva,Schroder:2024gsi,Azzola:2026cwa,Mariotti:2024eoh,Azzola:2024pzq,Azzola:2026cwa},
generate pulsar quakes or variations of the Earth's rotation period
\cite{Stadnik:2014cea},
or even explain the \emph{planes of satellites} galactic issue
\cite{Naik:2022lcn}.
A crucial consequence of a DW network evolution is the 
generation of a stochastic gravitational-wave background
(SGWB) \cite{Chang:1998tb,Gleiser:1998na,Hiramatsu:2010yz,Kawasaki:2011vv,Saikawa:2017hiv,Hiramatsu:2013qaa,Roshan:2024qnv,Blasi:2025tmn},
eventually accompanied by particle dark matter genesis
\cite{King:2023ztb,King:2023ayw,Borah:2024kfn}
and primordial black hole genesis
\cite{Ferrer:2018uiu,Gelmini:2022nim,Gelmini:2023ngs,Dunsky:2024zdo,Gouttenoire:2023gbn,Gouttenoire:2023ftk,Ferreira:2024eru,Gouttenoire:2025ofv}.

In particular, gravitational radiation is continuously emitted during 
the linear scaling regime.
In general, the largest gravitational emission occurs during the latest 
stages of DW evolution, when the DW-to-background energy density ratio is maximal. 
The resulting GW signal can span a broad range of frequencies and is 
actively searched for in pulsar timing arrays (PTA) and ground-based 
interferometers such as LIGO--Virgo--KAGRA, and is of strong interest 
for next-generation detectors including the Einstein Telescope (ET) and 
space-based missions such as LISA.

Over the last decades, increasing attention has been 
devoted to a complementary observational channel. 
In particular, if DWs (safely) persist to 
late times, their passage through the Earth's geospace may leave transient 
imprints on distributed networks of precision sensors. 
These include magnetometer networks for axion-based walls, particularly the GNOME project
\cite{Pospelov:2012mt,Pustelny:2013vza,Afach:2018eze,Masia-Roig:2019hsy,GNOME:2023rpz},
gravimetry
\cite{McNally:2019lcg},
terrestrial interferometers
\cite{Jaeckel:2020mqa,Alvarez:2025frb,Heisenberg:2023urf,Thoss:2025yht},
ground-based networks of optical atomic clocks
\cite{Wcislo:2016,Wcislo:2018ojh,Roberts:2019sfo},
or space-based networks of atomic clocks such as those onboard Global Navigation Satellite System (GNSS) constellations
\cite{Derevianko:2013oaa,Roberts:2017hla,Roberts:2018xqn,Roberts:2018agv,Panelli:2019qtl,Bertrand:2024ypu}.
In particular, a DW crossing would induce correlated 
signatures across spatially separated nodes of the network, producing a distinctive 
pattern that can be exploited as a detection strategy.
The features of the pattern depend on the coupling of the scalar field comprising the wall
with ordinary matter as well as on the transverse size of the topological defect.

In this paper, the impact of scalar field domain walls with characteristic sizes comparable to the Earth's radius
on the Galileo GNSS satellite constellation is studied. 
Approximately three years of satellite atomic clock data 
are used to test and simulate possible domain wall crossing events.

This paper, as part of the G4S$\_2.0$ research activities, 
investigates the impact of scalar field domain walls with characteristic sizes comparable to the Earth's radius on the Galileo GNSS satellite constellation. 
Approximately three years of satellite atomic clock data are used to search for and simulate possible domain wall crossing events.

The paper is organized as follows.
In Sec.~\ref{Early universe, ultralight scalar field and domain wall solution}, the adopted cosmological and particle physics model is described, the domain wall solution is reviewed, 
and a suitable coupling between the scalar field and selected sectors of the Standard Model is introduced -- 
together with its phenomenological consequences.
In Sec.~\ref{Domain wall dark matter and GNSS atomic clocks}, the impact of an Earth-sized domain wall on a GNSS constellation is discussed, 
focusing in particular on the data from the atomic clocks onboard the satellites.
The properties of the data and their impact on a possible physical signal are also described.
In Sec.~\ref{Domain walls and Galileo GNSS atomic clock noise}, 
the statistical procedures adopted to characterize the properties of the clock network are discussed,
\textit{namely} its capability to generate false signals and its efficiency in detecting real signals.
Finally, Sec.~\ref{Conclusions and prospects} concludes the work, summarizing the results and forthcoming developments
as well as discussing complementary aspects.

In this manuscript, where not explicitely stated otherwise,
\emph{natural units} of particle physics ($\hbar = c = k_B = 1$) are used
and the reduced Planck mass is defined as $M_\text{P} = 1/\sqrt{8\pi G_\text{N}}$ with
$G_\text{N}$ denoting Newton’s gravitational constant.
In addition, the \textit{mostly minus} Lorentzian metric signature is adopted, \textit{i.e.} $(+,-,-,-)$.


\section{Early universe, dark ultralight scalar field and domain walls}
\label{Early universe, ultralight scalar field and domain wall solution}

Suppose that the early universe, just below the epoch of spontaneous symmetry breaking (SSB) of the Standard Model (SM), 
is governed by a cosmological action of the form:
\begin{align}
    S[g_{\mu\nu},\phi,...] 
    &\sim S_{\text{EH}} + S_{\text{M}} \nonumber\\
    &= \int_{\mathcal{M}} \! d^4x\, \sqrt{-g}\,
    \Bigg\{ -\frac{M_P^2}{2}R + \mathcal{L}_{\text{SM}} + \mathcal{L}(\phi) \Bigg\},
\label{1}
\end{align}
where the first term, $S_{\text{EH}}$, represents the Einstein-Hilbert action for gravity with $R$ being the Ricci scalar curvature 
and $g$ the determinant of an almost flat Friedmann-Lema\^{i}tre-Robertson-Walker (FLRW) metric tensor $g_{\mu\nu}$. 
The second term, $S_{\text{M}}$, describes the matter content of the broken phase of the Standard Model 
$SU(3)_{\text{C}}\times U(1)_{\text{EM}}$, 
supplemented by a minimally coupled and canonically normalized scalar singlet $\phi$.
This field is thought to be neutral under the SM gauge symmetry
and charged under a new -- eventually approximate -- discrete symmetry $\mathbb{Z}_2$ which
does not affect the SM particles.

Within this framework, the scalar field $\phi$ is assumed to satisfy two key conditions. 
First, $\phi$ is posited to have a temperature-dependent potential $V(\phi,T)$ such that, 
as the universe expands and cools below some high critical temperature ($T\ll T_c$), 
it undergoes a smooth second-order phase transition. 
This transition leads to the spontaneous symmetry breaking (SSB) of the internal symmetry and 
generates a network 
of (eventually metastable) domain walls (DWs) throughout the universe, via the well-known Kibble-Zurek mechanism.
Second, $\phi$ is allowed to quadratically couple to Standard Model components through appropriate operators or \textit{portals},
which do not significantly alter or spoil the dynamics of the phase
transition and the final production of domain walls.

\subsection{Scalar field domain wall solution}
\label{Scalar field domain wall solution}

Considering the scalar field, one can write down the low energy lagrangian density as 
$\mathcal{L}(\phi) = \frac{1}{2}\partial_{\mu}\phi\partial^{\mu}\phi - V(\phi)$
where $V(\phi)$ represents the low or zero-temperature potential that can be approximated as
\begin{equation}
    V(\phi) \sim \frac{\lambda_{\phi}}{4}\left(\phi^2 - v^2\right)^2 + ...
\label{2}
\end{equation} 
where hypothetical $\mathbb{Z}_2$-breaking terms are supposed to be negligible at this stage. 
Here, the parameter $\lambda_{\phi}$ denotes the dimensionless (quartic) self-coupling of the scalar field 
and $|v|$ is the (absolute) vacuum expectation value (vev) characterizing the phase transition and the vacuum manifold's degenerate minima ($\pm v$).
In the Minkowski spacetime limit, the solution of the corresponding Bogomolny-Cauchy problem for a 
scalar field solely depending on the $x$-direction
gives rise to the following non-perturbative, topological, and Lorentz-invariant (kink) field profile
\begin{equation}
    \phi(x) = v\tanh\left(\frac{x}{d}\right), \quad d = \sqrt{\frac{2}{\lambda_{\phi}}}\frac{1}{v}
\label{3}
\end{equation}
which represents a (locally) planar domain wall configuration
interpolating between the two minima $\pm v$.
The center of the topological defect ($\phi=0$) preserves the original symmetry
and represents the relic of the early phase.
In this scenario, the \textit{cosmological embedded} domain wall $\phi(x,y,z)$ 
lies perpendicular to the $x$-axis and extends infinitely in the 
$(y,z)$-plane, given a suitable reference frame. 
The domain wall curvature eventually manifests only on extremely large lengths, the horizon scale for instance.
The parameter $d$ describes the thickness of the domain wall, \textit{i.e.} the region of space 
where $d\phi/dx$ is not negligible (a sort of DW \emph{core}),
and is related to the \textit{reduce} Compton wavelength as:
\begin{equation}
    d = 2\lambda_{\rm Compton}=\frac{2}{m_{\phi}}, \quad m_{\phi} = \sqrt{2\lambda_{\phi}}v,
\label{4}
\end{equation}
where $m_{\phi}$ is the fundamental mass of the scalar field quanta,
usually inferable from the expansion of the scalar potential around (one of) the minima.
The case of ultralight scalar fields with masses $m_{\phi} \lesssim  10^{-10}$ eV 
-- hence, much smaller than electron or expected SM neutrino masses --
typically leads to DWs with large lifetime and large transverse size $d$,
for example, $\geq 10^2 - 10^3$ km.

The domain wall energy configuration is given by the computation of the corresponding stress energy tensor $T_{\mu\nu}$.
In particular, the \textit{surface energy density}, \textit{i.e.} the time-time component $T_{00}$, reads
\begin{equation}
    \rho_{\phi}(x) = \frac{v^2}{d^2} \frac{1}{\cosh^4(x/d)},
\label{5}
\end{equation}
which, near the relic, unbroken phase at $x\sim 0$, can be perturbatively expanded as:
\begin{equation}
    \rho_{\phi}(x) \sim \left(\frac{v}{d}\right)^2 \left[ 1 - 2\left(\frac{x}{d}\right)^2 + \frac{7}{3}\left(\frac{x}{d}\right)^4 + \dots\right].
\label{6}
\end{equation}

The spatial integration of the surface energy density provides the energy per unit area, namely the \textit{surface energy}, 
\begin{equation}
    \sigma_{\phi} = \frac{4v^2}{3d} \, \simeq \sqrt{\lambda_{\phi}} v^3.
\label{7}
\end{equation}

\begin{figure*}
    \centering
    \includegraphics[
      width=1\textwidth,
    ]{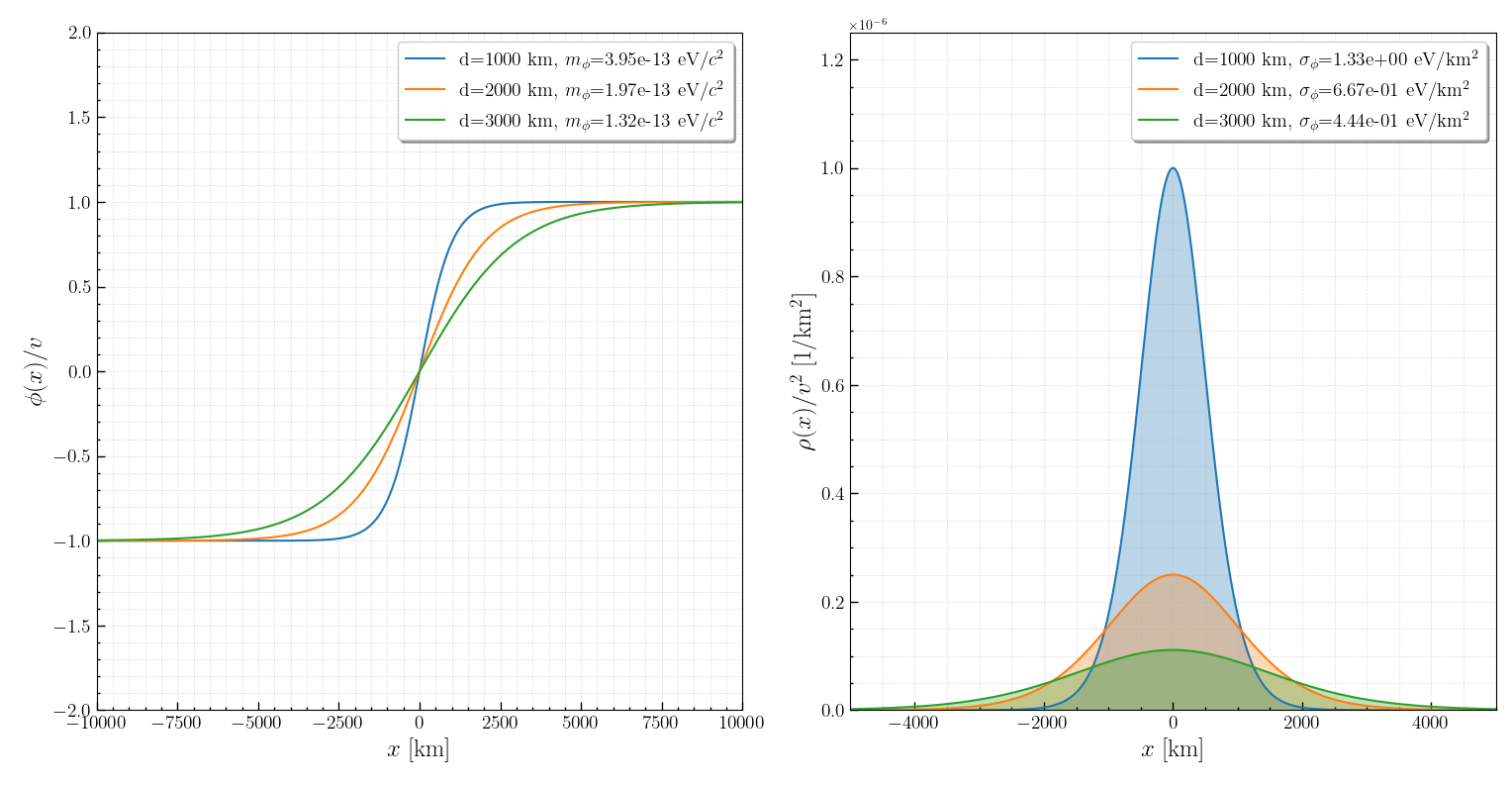}
    \caption{\it Left panel: DW field space profile for three different sizes.
    The normalized field space solution gets steeper and steeper as the mass of the scalar field becomes larger.
    Right panel: DW energy density space profile for the same sizes and $v^2\sim 10^3$ \text{eV/km}.
    The energy density sharpens with decreasing $d$, increasing the tension and approximating a single-particle behavior.
    The plot is shown SI standard units.}
    \label{Fig1}

    \centering
    \includegraphics[
      width=1\textwidth,
    ]{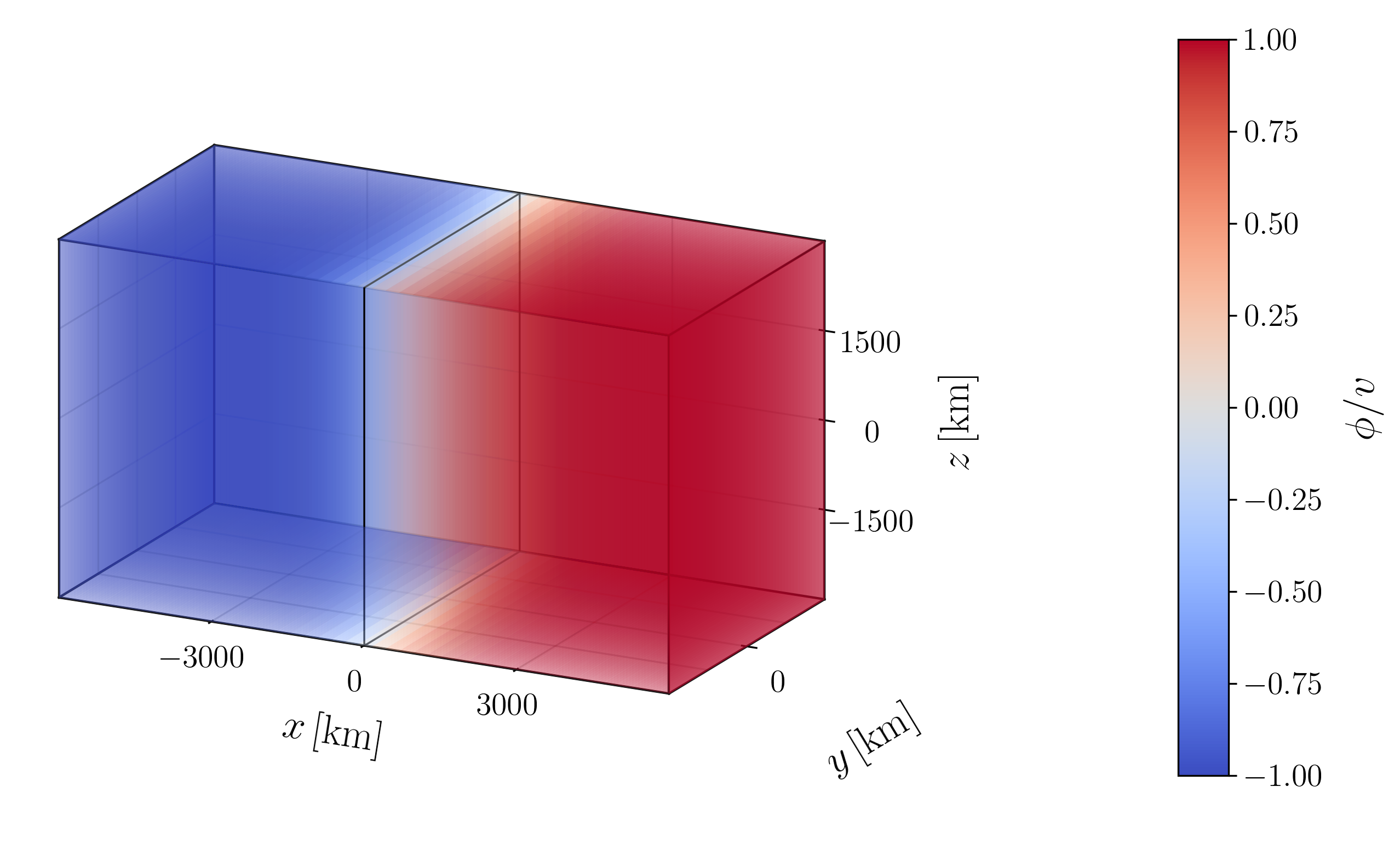}
    \caption{\it The three dimensional embedding of the normalized topological domain wall solution in a finite region of space. 
    The thickness of the wall is assumed to be $d\sim 10^3$ km. The white region across $x\sim 0$ indicates that the scalar field
    gradient $d\phi/dx$ experiences a strong variation. The plot is shown SI standard units.}
    \label{Fig2}
\end{figure*}

It is worth noticing that the $x$-component of the pressure vanishes, \textit{i.e.}, $T_{xx} = 0$, 
while the transverse components satisfy $T_{yy} = T_{zz} = -\rho_{\phi}$.
As a consequence the stress energy tensor can be written in compact form as $T_{\mu\nu} = \rho_{\phi}(x)(1,0,-1,-1)$
and the integral of the pressure component(s) provide a (negative) \textit{surface tension}, 
described by Eq.~\eqref{7}.
Fig.~\ref{Fig1} illustrates numerical examples of DW field solutions and energy densities for
ultralight scalar masses, while Fig.~\ref{Fig2} shows the three dimensional embedding of the solution.

The distribution of domain walls
is usually described by the average energy density 
\begin{equation}
    \rho_{\text{walls}} = \frac{\sigma_{\phi}}{L}
\label{8}
\end{equation}
where $L$ is the \textit{characteristic length} of the network, describing the (statistical) average distance of a couple of \textit{close} domain walls.
In general, the evolution of $L$ with cosmic time is slow compared to typical human timescales, 
so both $L$ and $\rho_{\text{walls}}$ can be regarded as nearly constant to date.
The energy density of the network Eq.~\eqref{8} can also be expressed
in terms of the single domain wall size as
\begin{equation}
    \rho_{\text{walls}}\sim \rho_{\text{inside}}\frac{d}{L}, \quad \rho_{\text{inside}} \simeq \frac{v^2}{d^2}
\label{9}
\end{equation}
where $\rho_{\text{inside}}$ is the leading term of Eq.~\eqref{6}.
The surface energy $\sigma_{\phi}$ also parameterizes the time scale for a hypothetical domain wall dominance as $t_{\text{dom}} \sim 3M^2_{\text{P}} / 4A\sigma_{\phi}$, where $A$ is a constant (numerically estimated to be $\mathcal{O}(1)$). The absence of this phase in the observable universe leads to $\sigma_{\phi} \ll 2 \times 10^{-6}$~GeV$^3$, 
which reduces to a bound on the scalar field vev $v^3 \ll (10^{-2}~\text{GeV})^3$ (ZKO bound) in the limit of $\lambda_{\phi} \sim 1$ ~\cite{Zeldovich:1974uw,Vachaspati:2006zz}.

\subsection{Scalar field domain wall coupling to matter}
\label{Scalar field domain wall coupling to matter}

The interaction between the scalar field and SM here
is conceived just by a simple scalar-QED portal 
and modeled through Lorentz and gauge-invariant couplings. 
In the context of the effective field model building, 
various interaction terms can be introduced, 
such as scalar-fermion ($\sim\phi^n\bar{\psi}\psi$) or scalar-tensor ($\sim\phi^m |F_{\mu\nu}|^2$) couplings.  
Here, the focus is on non-renormalizable quadratic couplings in the dark scalar field $\phi$, with no gravity-mediated effects
\begin{align}
    \mathcal{L}_{\phi,\text{QED}} 
    &= \bar{\psi_e}\left(i\gamma^{\mu}D_{\mu}-m_e\right)\psi_e
    - \frac{1}{4} |F_{\mu\nu}|^2 \nonumber\\
    &\quad - g_{\phi e}(\phi^2-v^2)\bar{\psi_e}\psi_e \nonumber\\
    &\quad + \frac{g_{\phi \gamma}}{4}(\phi^2-v^2)|F_{\mu\nu}|^2.
\label{10}
\end{align}
Here, $\psi_e$ is the electron Dirac (spinor) field, $\bar{\psi}_e$ its adjoint, 
$m_e$ is the electron mass, $\gamma^{\mu}$ are the Dirac matrices, 
$D_{\mu} = \partial_{\mu} - ieA_{\mu}$ is the covariant derivative (where $e$ is the electron charge)
and $F_{\mu\nu}=\partial_{\mu}A_{\nu} - \partial_{\nu}A_{\mu}$ is the field strength of the photon field $A_{\mu}$. 
The constants $g_{\phi e}$ and $g_{\phi \gamma}$ parameterize the interaction strength with electrons and photons, respectively, 
and can be expressed in terms of effective energy scales $\Lambda_e$ and $\Lambda_{\gamma}$ as:
\begin{equation}
    g_{\phi e} = \frac{m_e}{\Lambda^2_e}, \quad g_{\phi\gamma} = \frac{1}{\Lambda^2_{\gamma}}.
\label{11}
\end{equation}

The quantities $\Lambda_e$ and $\Lambda_{\gamma}$ describe the range of the validity of the effective description
and are expected to be much larger than the vacuum expectation value $v$ of the dark field. 
As mentioned in Sec.\ref{introduction}, one assumes that the coupling terms in Eq.~\eqref{10}
do not alter the dynamics of the early universe thermal phase transition:
the critical temperature can be above the Big Bang Nucleosynthesis scale ($T_{\text{BBN}} \sim 1 $ MeV)
or even the QCD transition ($T_{\text{QCD}}\lesssim 200$ MeV)
while the formation of domain walls proceeds in the usual way.
The details of the phase transition and its impact
on the production of the target (observable) walls will be discussed in a dedicated work.

If a region of ordinary matter crosses a DW,
the lagrangian density in Eq.~\eqref{10} leads to temporary variations of the electron mass $m_e$ and the 
coupling strength of the electromagnetism, the fine structure constant $\alpha$
\begin{align}
    m_{e,\text{eff}}(\phi)
    &= m_e\left[1 + \frac{\phi^2-v^2}{\Lambda_e^2}\right],
\label{12}\\
    \alpha_{\text{eff}}(\phi)
    &= \frac{\alpha}{\left[1 - \frac{\phi^2-v^2}{\Lambda_{\gamma}^2}\right]}
    \simeq \alpha\left[1 + \frac{\phi^2-v^2}{\Lambda_{\gamma}^2}\right],
\label{13}
\end{align}
allowing the definition of corresponding relative variations:
\begin{align}
    \frac{\Delta m_e}{m_e} 
    &= \frac{\phi^2-v^2}{\Lambda_e^2}, 
    \quad \Gamma_e=\Lambda_e^{-2},
\label{14}\\
    \frac{\Delta\alpha}{\alpha} 
    &\simeq \frac{\phi^2-v^2}{\Lambda_{\gamma}^2}, 
    \quad \Gamma_{\gamma}=\Lambda_{\gamma}^{-2},
\label{15}
\end{align}
where $\Gamma_e$ and $\Gamma_{\gamma}$ often compare in literature as \textit{effective couplings}, 
see for instance \cite{Derevianko:2013oaa,Roberts:2017hla}. 
Note that, Eq.~\eqref{13} is derived by assuming that the coefficient of the 
Maxwell kinetic term can be reworded as $e^2/16\pi\alpha$, where $e^2=4\pi\alpha$ holds in natural units.
Moreover, the cut-off energy scale or the effective strength can be compared in literature 
as $\Lambda_{\alpha}$ and $\Gamma_{\alpha}$, respectively, to indicate that they are closely related
to the fine-structure constant physics.

\begin{figure}
    \centering
    \includegraphics[
      width=0.5\textwidth,
    ]{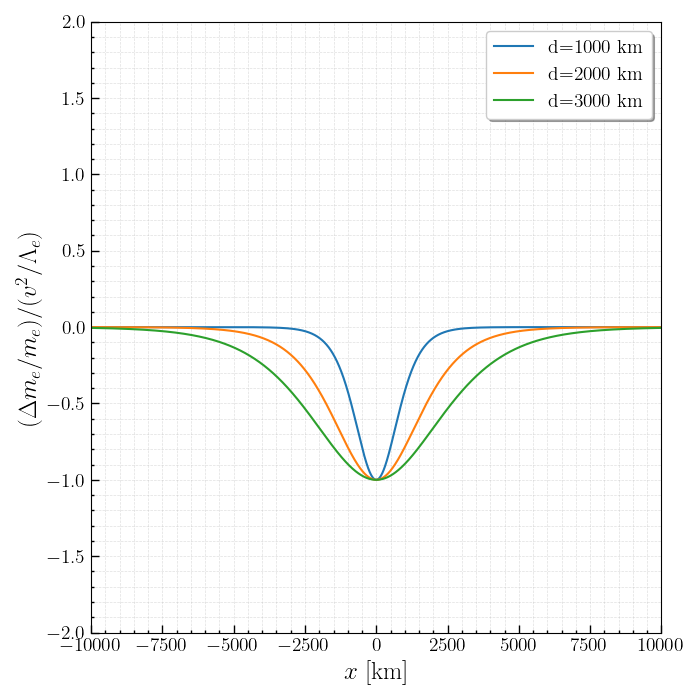}
    \caption{\it Normalized electron mass variations for three DW sizes in SI units. 
    The negative, nearly gaussian variation arises from the dark scalar-electron portal. The spike's features depend on DW size and the vev-to-(operator) energy scale ratio.}
    \label{Fig3}
\end{figure}

In the simplest case, one can suppose a crossing event occurring along the $x$-axis.
In this scenario, the fundamental constants recover their original values far from the DW ($|x|\gg d$) while
resemble negative spikes with features controlled by the size $d$ and the ratio $v^2/\Lambda_X^2, X=e,\gamma$
approaching the scales of the DW core ($|x|\sim d$).
This means that a hypothetical fundamental constant measurement inside the DW
provides a nonstandard result because the original vacuum symmetry is restored,
corresponding to a temporary local position invariance (LPI) violation.
The specific result will depend on the space(time) position at which the measurement is performed.
Fig.~\ref{Fig3} shows the normalized electron mass variation for different DW sizes.
As the domain wall thickness increases, the variation becomes less sharped.

In general, one can also assume phenomenological \textit{wrong-sign} couplings:
\begin{equation}
    \mathcal{L}_{\text{int}} = g_{\phi e}(\phi^2-v^2)\bar{\psi_e}\psi_e - \frac{g_{\phi \gamma}}{4}(\phi^2-v^2)|F_{\mu\nu}|^2,
\label{16}
\end{equation}
leading to relative fundamental constants variation of the form
\begin{align}
    \frac{\Delta m_e}{m_e} 
    &= -\frac{\phi^2-v^2}{\Lambda_e^2},
\label{17}\\
    \frac{\Delta\alpha}{\alpha} 
    &\simeq -\frac{\phi^2-v^2}{\Lambda_{\gamma}^2}.
\label{18}
\end{align}
These variations are characterized by a positive spike near the DW center
which is consistent with the phenomenological gaussian \textit{ansatz} for the scalar field
(preliminary) proposed by \textit{Derevianko and Pospelov} \cite{Derevianko:2013oaa} or \textit{Roberts et al.} \cite{Roberts:2017hla}.
Inhomogeneous-sign couplings which induce, for example, a negative spike for the electron mass and a positive one
for the fine structure constant (or vice versa), are also possible.
However, it is crucial to stress that the viability of these scenarios 
is strictly connected with the scale of the high energy physics.
It is worth mentioning that an equivalent analysis could also be performed for the opposite field solution (the antikink)
with the same, general results.

\section{Domain walls, dark matter and GNSS atomic clocks}
\label{Domain wall dark matter and GNSS atomic clocks}

\subsection{Basic theory}
\label{Basic theory}

The hypothesized distribution of scalar field domain walls, according to some,
may contribute to the local or (Milky Way) galactic dark matter energy density
and could be described via the standard Halo Model for dark matter, see
\cite{Derevianko:2013oaa,Roberts:2017hla} or \cite{Bertrand:2024ypu} for details.
The detection of these domain walls
can be pursued analogously to gravitational wave searches, for instance, 
through a dedicated and correlated \textit{distributed system} of advanced quantum devices\footnote{In this context, quantum sensors are understood as devices whose operation relies on quantum physics, even though their final outputs may be classical (\textit{e.g.}, atomic clocks) or quantum (\textit{e.g.}, interferometers).}.

In this context, the existing GNSS infrastructures would represent a compelling opportunity.
Indeed, GNSS are basically constituted by 
\textit{(i)} a cloud of geospatial satellites orbiting Earth, each equipped with a pair of advanced space-based atomic clocks (a primary and a secondary or auxiliary clock) and
\textit{(ii)} a network of ground-based reference stations hosting atomic clocks that serve as time-reference for the satellite clocks.

The atomic clocks used for GNSS missions are \textit{microwave} clocks, \textit{i.e.}, Caesium, Hydrogen and Rubidium, 
and have compact sizes and weights.
In general, an atomic clock is naturally characterized by a nominal working frequency $\nu_0$ that depends on the nature of the clock. 
In particular, this frequency depends on the energy gap $\Delta E$ between the two atomic levels 
of the hyperfine transition characterizing the species, see \cite{Lewis:1991FrequencyStandards}
and Table \ref{Tab1} for details.

\begin{table*}[t]
\centering
\small
\renewcommand{\arraystretch}{1.25}

\begin{tabular*}{\textwidth}{@{\extracolsep{\fill}} l l l @{}}
\hline\hline

\parbox[t]{0.22\textwidth}{Clock type}
& \parbox[t]{0.47\textwidth}{Electronic state $n\,^{2s+1}\mathrm{L}_J$ and selected hyperfine transition}
& \parbox[t]{0.22\textwidth}{Nominal frequency $\nu_0$ [Hz]} \\
\hline

\parbox[t]{0.22\textwidth}{Caesium clock, $^{133}\mathrm{Cs}$}
& \parbox[t]{0.47\textwidth}{$6\,^2\mathrm{S}_{1/2}$, $|F=3,m_F=0\rangle\leftrightarrow|F=4,m_F=0\rangle$}
& \parbox[t]{0.22\textwidth}{$9\,192\,631\,770$} \\

\parbox[t]{0.22\textwidth}{Hydrogen Maser, $^{1}\mathrm{H}$}
& \parbox[t]{0.47\textwidth}{$1\,^2\mathrm{S}_{1/2}$, $|F=0,m_F=0\rangle\leftrightarrow|F=1,m_F=0\rangle$}
& \parbox[t]{0.22\textwidth}{$1\,420\,405\,751.768$} \\

\parbox[t]{0.22\textwidth}{Rubidium clock, $^{87}\mathrm{Rb}$}
& \parbox[t]{0.47\textwidth}{$5\,^2\mathrm{S}_{1/2}$, $|F=1,m_F=0\rangle\leftrightarrow|F=2,m_F=0\rangle$}
& \parbox[t]{0.22\textwidth}{$6\,834\,682\,610.90429$} \\

\hline\hline
\end{tabular*}

\vspace{0.15cm}

\parbox{\textwidth}{
\caption{\it
Relevant atomic transitions for GNSS atomic frequency standards.
The notation $n\,^{2s+1}\mathrm{L}_\mathrm{J}$ labels the electronic state in which the hyperfine transition occurs.
Here $n$ denotes the principal electronic quantum number, 
$2s+1$ the spin multiplicity, with $s$ the total electronic spin, 
$\mathrm{L}$ the electronic orbital angular momentum, and $\mathrm{J}$ the total electronic angular momentum 
($\mathbf{J}=\mathbf{L}+\mathbf{S}_{\rm e}$, with $\mathbf{S}_{\rm e}$ the total electronic spin angular momentum).
All transitions listed here occur in states with $\mathrm{L} = 0$, conventionally denoted by the spectroscopic letter $\mathrm{S}$.
The hyperfine states are specified by $F$ and $m_F$. The first represents the total atomic angular momentum quantum number while the second the related projection along the quantization axis.
In particular, the choice $m_F=0$ provides state(s) with a very good frequency stability, 
since these transitions are typically weakly sensitive to magnetic-field fluctuations.
The hydrogen maser technology 
can be implemented in either active (AHM) or passive (PHM) configurations (the latter being used on-board Galileo satellites).
Their operating principles are different but the underlying atomic physics and the final reference frequency is the same.
}
\label{Tab1}
}

\end{table*}

This frequency is a function of some fundamental constants like electron mass, hadron masses or fine structure constant
with the details depending on the specific atomic scenario.
In particular, the Caesium-133 (Cs) atomic clock (master clock on-board of GPS satellites) has the nominal frequency $\nu_0\sim 9.2$ GHz, 
the Passive Hydrogen Maser (PHM) (master clock on-board Galileo satellites) has a frequency $\nu_0\sim 1.4$ GHz
while for the Rubidium-87 clock (RAFS) (used as a secondary clock on both GPS and Galileo spacecrafts) one has $\nu_0\sim 6.8$ GHz,
see Table \ref{Tab1}.
In general, an atomic clock can be considered as a localized region of ordinary matter sensitive to variations in the physical environment.
In light of this, if a GNSS satellite or reference atomic clock traversed a scalar field DW configuration 
it would experience a variation of (some) fundamental constants that could perturb 
the quantum physics of the hyperfine transition
and, consequently, the frequency from the nominal value $\nu_0$ to a new one $\nu$.
In particular, the dimensionless \textit{fractional frequency} (FF) variation can be expressed as a linear combination of the relative fundamental constants variations.
Under the scalar field-QED coupling assumed in Sec.~\ref{Early universe, ultralight scalar field and domain wall solution}, 
one obtains
\begin{equation}
    \frac{\Delta \nu}{\nu_0} = \sum_X K_X \frac{\Delta X}{X} = K_e \frac{\Delta m_e}{m_e} + K_{\alpha}\, \frac{\Delta \alpha}{\alpha}
\label{19}
\end{equation}
where $\Delta \nu = \nu - \nu_0$ and 
the $K_X, X=e,\alpha$ are the so-called \textit{sensitivity coefficients} describing the importance of the associated fundamental variations.
Some comments are in order.
First, the value of the sensitivity coefficients strongly depends on the type of atomic clock.
For example, in the case of a Cs clock
one has $K_e=2$ and $K_{\text{$\alpha$}} = 4.83$,
while for a PHM one has $K_e=2$ and $K_{\text{$\alpha$}} = 4$
and for a RAFS, $K_e=2$ and $K_{\text{$\alpha$}} = 4.34$.
Second, (along the $x$ direction) the FF variation Eq.~\eqref{19} depends on the magnitude of the microphysics parameters and on the sign of the assumed coupling terms in the interaction lagrangian density.
For example, in the case of Eq.~\eqref{10} or Eq.~\eqref{16} the two contributions in Eq.~\eqref{19} \textit{positively interfere}
while in the other cases they can \textit{negatively interfere} with each other.

The occurrence of the FF variation tends to perturb the \textit{nominal} time reading 
of the considered atomic clock.
In particular, the \textit{accumulated} time reading or \textit{intrinsic bias} of the clock
can be computed 
as the integral over the fractional frequency variation
\begin{equation}
    \Delta t = \int_{\mathcal{I}} \! dt'\, \frac{\Delta \nu(t')}{\nu_0}
\label{20}
\end{equation}
where $\mathcal{I}$ represents the integration interval related to the global scalar DW-clock interaction time 
and, in principle, $\mathcal{I}=(-\infty,+\infty)$.

In GNSS missions, the evolution of a single clock's FF variation cannot be directly observed.  
However, it can be observed using a ground clock as a reference: the link between a satellite clock and a given ground station clock allows one to
construct and deploy an alternative type of bias, \textit{i.e.} the so-called \textit{GNSS (clock) bias} or simply \textit{bias}.
The GNSS bias describes the \textit{time displacement} of the satellite clock with respect to a reference clock, \textit{i.e.}
\begin{equation}
    S(t) = {T}_{\text{sat}}(t) - {T}_{\text{ref}}(t),
\label{21}
\end{equation}
where $t$ denotes a continuous time variable, 
${T}_{\text{sat}}(t)$ is the time reading of the satellite clock, 
and ${T}_{\text{ref}}(t)$ is the time reading of the reference clock.
In the ideal case, the bias is a constant quantity: it is positive if 
${T}_{\text{sat}}>{T}_{\text{ref}}$, negative if 
${T}_{\text{sat}}<{T}_{\text{ref}}$, and practically zero if the two clocks are perfectly aligned for all $t$.

Note that, 
different clocks operate in different periods to produce a global bias time series associated with a given satellite.
Thus, one portion of the time series may be associated with a given onboard clock, 
whereas a subsequent portion may be associated with another one, 
depending on which clock is chosen as the main one onboard the satellite during some time intervals.

The passage of a domain wall can trigger a bias evolution for a given satellite clock 
that can be written as
\begin{equation}
    S(t) = \int_{-\infty}^{t} \! dt'\, \left[\left(\frac{\Delta \nu(t')}{\nu_0}\right)_{\text{sat}} - \left(\frac{\Delta \nu(t')}{\nu_0}\right)_{\text{ref}}\right]
\label{22}
\end{equation}
with the integrand representing a (GNSS) \textit{rescaled} fractional frequency (RFF), attributable to each satellite (clock).
In the scalar-QED interacting scenario, the bias assumes the form
\begin{align}
    S(t) 
    &= \int_{-\infty}^{t} \! dt'\, \Bigg\{
    v^2\Gamma_{\text{eff,sat}}
    \left[\tanh^2\left(\frac{x_{\text{sat}}}{d}\right) - 1\right] \nonumber\\
    &\quad - v^2\Gamma_{\text{eff,ref}}
    \left[\tanh^2\left(\frac{x_{\text{ref}}}{d}\right) - 1\right]\Bigg\}.
\label{23}
\end{align}
where the global effective couplings are
\begin{equation}
    \Gamma_{\text{eff,i}} = K_{e,i}\Gamma_e + K_{\alpha,i}\Gamma_{\gamma}, \quad i=\text{sat, ref}
\label{24}
\end{equation}
and simplified laws of motion are assumed
\begin{align}
    x_{\text{ref}} 
    &\sim v_g(t'-t_{\text{ref}}),
\label{25}\\
    x_{\text{sat}} 
    &\sim v_g(t'-t_{\text{sat}}).
\label{26}
\end{align}
Here, $v_g$ is the speed at which a GNSS satellite or station should traverse a domain wall,
\textit{i.e.} the relative galactic velocity of the DW due to Earth motion in the Milky Way.
This parameter can be statistically described via a proper Maxwell distribution, following the assumed Halo Model for the domain wall dark matter \cite{Roberts:2017hla}, and typically $v_g\sim 270$ km/s\footnote{In principle, a satellite crossing a domain wall could experience an effective wall thickness $d_{\mathrm{eff}}$ -- with $d_{\mathrm{eff}} = d/\gamma_l$, where $\gamma_l$ is the Lorentz factor -- due to relativistic length contraction. 
As a result the final clock bias could be affected. 
However, such effects are negligible for the typical relative velocities ($v_g \sim 10^{-3}c$) expected in the Galactic halo.}.
At the same time, $t_{\text{sat}}$ represents the interaction epoch of the satellite clock with the DW center while $t_{\text{ref}}$ represents the corresponding one for the reference clock.
As can be seen, the global function $S(t)$ in Eq.~\eqref{23} is generated by the integration of two temporal and consecutive contributions.

Supposing $t_{\text{sat}}<t_{\text{ref}}$, the first contribution describes a first negative RFF spike
\textit{directly} experienced by the satellite clock due to its interaction with the DW
while the second contribution determines a positive RFF spike \textit{indirectly} experienced by the satellite clock due to the 
DW interaction with the ground station clock.
As a consequence, the amplitude of the first spike depends on the type of the active satellite clock,
while the amplitude of the second spike on the properties of the reference clock.

\begin{figure*}
    \centering
    \includegraphics[
      width=1\textwidth,
    ]{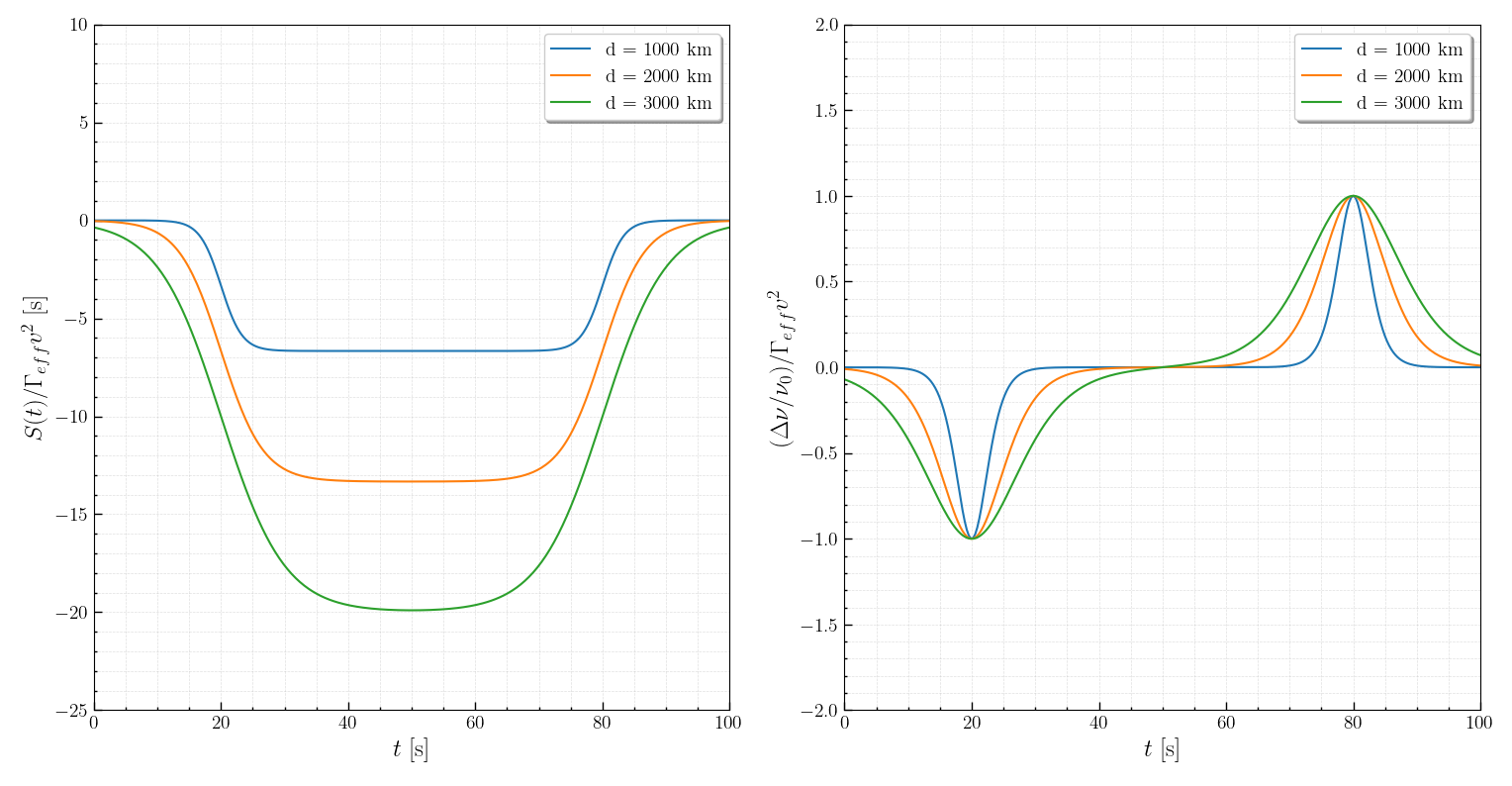}
    \caption{\it Left panel: Normalized clock bias solution for three different DW sizes and with interaction epochs with the DW center
    $t_{\text{sat}}=20$s and $t_{\text{sat}}=80$s. 
    The clock bias shows an almost boxcar behavior.
    The first drop in the bias describes the DW-satellite clock interaction and the related misalignment of the satellite clock 
    with respect to the reference one.
    As the DW moves toward the reference clock the bias is set to a new (negative) value.
    Finally, as the DW reaches the reference clock the bias shows an increase that indicates the realignment.
    The features of the signal are controlled both by the fundamental physics of the DW-electron interaction 
    and by the DW velocity, or equivalently the average time $\tau_{\text{avg}}$.
    In principle, larger sizes naturally induce larger bias excursions.
    Right panel: Normalized RFF evolution for the same DW sizes and interaction epochs. 
    The function shows two spike-like signals at the interaction epochs.
    The amplitude of the signal depends on the clock type and the scalar coupling while the shape depends on the DW size.}
    \label{Fig4}
\end{figure*}

The bias definition and the expected inhomogeneous response of satellite clocks
make it impossible to compute a unique, simple and explicit analytical form for $S(t)$.
Therefore, one can consider a teaching \textit{case study}
adopting the following assumptions:
\begin{enumerate}[label=\textit{\arabic*})]
    \item A finite lower integration bound $a$ is introduced in place of $-\infty$, with the limit $a \to -\infty$ taken after integration. Alternatively, $a$ may be kept finite, since $S(t)$ describes local effects whose physical characterization does not require integration over the entire past history.
    \item Suppose that the scalar-electron coupling dominates the interaction ($\Lambda_e\ll\Lambda_{\gamma}$)
    allowing the same response for the main atomic clocks.
    In this way, it is not important if the active clock is the primary clock
    or the secondary one and $\Gamma_{\text{eff}}$ is reduced to $2\Gamma_e$.
    \item Select a ground station clock with similar properties.
\end{enumerate}
In this setup, the bias integral $S(t)$ becomes
\begin{align}
    S(t) 
    &\simeq v^2\Gamma_{\text{eff}}\int_{a}^{t} \! dt'\, \Bigg\{
    \tanh^2\left[\frac{v_g(t'-t_{\text{sat}})}{d}\right]\nonumber\\
    &\quad - \tanh^2\left[\frac{v_g(t'-t_{\text{ref}})}{d}\right] \Bigg\}.
\label{27}
\end{align}
The corresponding solution can be written as
\begin{align}
    S(t) 
    &\sim \frac{\Gamma_{\text{eff}}\, v^2\, d}{v_g}
    \Bigg\{ -\tanh\left[\frac{v_g(t'-t_{\text{sat}})}{d}\right] \nonumber\\
    &\qquad + \tanh\left[\frac{v_g(t'-t_{\text{ref}})}{d}\right] \Bigg\}.
\label{28}
\end{align}
where the ratio $d/v_g$ can be thought of as the \textit{average} or \textit{effective interaction time} $\tau_{\text{avg}}$ 
between the DW core region -- described by the thickness $d$ -- and the given clock.
This result indicates that the GNSS bias $S(t)$ acquires the form of an almost symmetric (negative) \textit{boxcar} function
whose basic features depend on the fundamental physics.
In particular, the time extension of the plateau $|t_{\text{ref}}-t_{\text{sat}}|$ keeps information about the relative galactic velocity of the topological defect to cover the 
distance between the satellite clock and the reference clock.
At the same time, the computation of the bias derivative $\dot{S}(t)$ provides the original, (directly) unobservable RFF.
In Fig.~\ref{Fig4} are shown some numerical examples for the resulting bias and RFF experienced by a satellite clock,
for some domain wall sizes, under the assumed recipe.

\begin{figure*}
    \centering
    \includegraphics[
      width=1\textwidth,
    ]{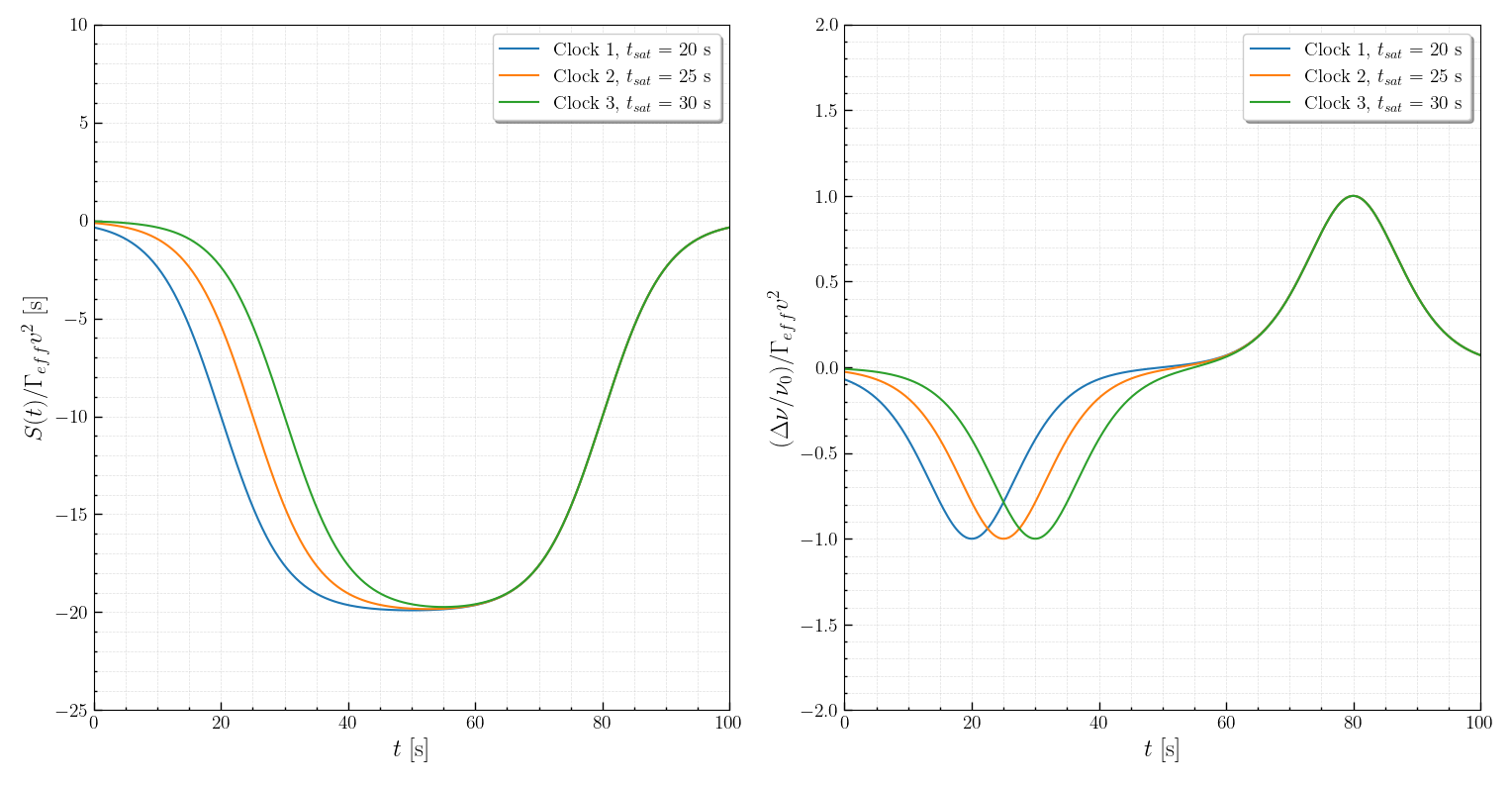}
    \caption{\it 
    Left panel: Normalized clock-bias pattern induced by a DW crossing in a very minimal network of three satellite clocks,
    with different satellite interaction epochs $t_{\rm sat},i$ and a common reference-clock epoch.
    The signals share the same amplitude but show different time extensions related to the DW propagation-time across the network.
    Right panel: Corresponding normalized RFF pattern, showing the satellite signals at different $t_{\rm sat},i$
    and the common one associated with the reference clock.
    }
    \label{Fig5}
\end{figure*}

The discussed framework can be ideally generalized by considering a network of $N$ satellite clocks and a selected ground reference clock.
In the limit described by Eq.~\eqref{28},
the DW generates a \textit{complete pattern} of bias(es) with $N$ simple boxcar features.
These signals have the same amplitude
but diverse time extensions $|t_{\text{ref}}-t_{\text{sat},i}|$.
These differences encode the time required by the DW to
move from the satellite clock $i$ to the ground station clock or viceversa. 
Furthermore, there is a pattern of RFF characterized by two basic properties.
First, all the RFF time series are expected to share a common positive spike
at the same epoch $t_{\text{ref}}$, due to the DW interaction with the ground station clock, \textit{i.e. Ground-Station Spike Event}
(GSSE).
The collection of all these single events determines a \textit{coincidence event}, namely, a trigger signal.
Second, all the RFF time series are also expected to show a distribution of negative spikes at times $t_{\text{sat},i}$ which are different, in principle, from satellite to satellite
and from $t_{\text{ref}}$.
Here, each single negative spike represents the interaction of the DW with a single satellite clock and can be referred to as 
a \textit{Satellite Spike Event} (SSE).
Finally, the distance $\tau_{\text{GNSS}}$ between the first and the last spike determines the time to cross the chosen clock network. 
Naturally, one should also expect a negligible bias variation (or RFF) for the clock $i$ if $t_{\text{ref}}\sim t_{\text{sat},i}$,
since the positive and negative contributions almost cancel each other out.
Fig.~\ref{Fig5} shows some examples of bias and RFF patterns, focusing on a simple set of just three clocks.

The wrong-sign coupling described by Eq.~\eqref{17} determines
a positive boxcar function while the corresponding RFF (with $t_{\text{sat}}<t_{\text{ref}}$) 
displays SSE/GSSE with switched sign.

The maximum amplitude of theoretical bias excursions for the standard and wrong-sign cases is obviously the same 
and can be easily estimated from Eq.~\eqref{28} as
\begin{equation}
    |S(t_{\text{sat}}<t<t_{\text{ref}})| \leq 2\Gamma_{\text{eff}}v^2 \frac{d}{v_g} = 4\frac{v^2}{\Lambda_e^2}\frac{d}{v_g} 
\label{29}
\end{equation}
In the case of $v^2/\Lambda_e^2\sim 10^{-12}$, $d\sim 10^3$ km and $v_g\sim 300$ km$/$s one has 
$\tau_\text{avg}\sim 3$ s,
$\tau_{\text{GNSS}}\sim 3 $ min and
$|S_{\text{max}}|\sim 4\times 10^{-11}$ s.
The amplitude of the signal induced in the RFF
near the interaction time (with satellite or station)
$t\sim t_{\text{int}}$ 
can be computed instead as
\begin{equation}
    |\dot{S}(t\sim t_{\text{int}})| \sim \frac{2v^2}{\Lambda_e^2}
\label{30}
\end{equation}
and basically depends on the underlying particle model.
Note that, the combination of Eqs.~\eqref{6} and \eqref{7} reveals that the vacuum expectation value of the scalar field
comprising the domain walls can be written as
\begin{equation}
v^2 = \frac{3}{4}\rho_{\text{walls}}\, d\, L, \quad L = v_g \mathcal{T}
\label{31}
\end{equation}
where $\mathcal{T}$ is the \textit{average encounter time} of the Earth between two consecutive domain walls of the 
distribution -- or in other words, the inverse of the expected encounter rate.
In light of this, the (absolute value of the) amplitude becomes 
\begin{equation}
|\dot{S}(t\sim t_{\text{int}})| \sim \frac{\rho_{\text{walls}}\, d\, v_g\, \mathcal{T}}{\Lambda^2_e}
\label{32}
\end{equation}
and the corresponding electron mass variation is 
\begin{equation}
\frac{\Delta m_e}{m_e} \sim \frac{\rho_{\text{walls}}\, d\, v_g\, \mathcal{T}}{K_e \Lambda^2_e}
\label{33}
\end{equation}

If one relaxes the computational conditions \textit{2)} and \textit{3)},  
the $\Delta\nu/\nu_0$ of the standard and wrong-sign cases provides
\textit{(i)} two \textit{asymmetric} bias boxcar functions -- one associated with the primary clock  
and the other with the secondary clock -- where the rising amplitude  
differs from the subsequent falling amplitude;  
\textit{(ii)} two types of SSE with different magnitudes,  
each associated with one of the two onboard atomic clocks and 
\textit{(iii)} GSSEs whose amplitude  
differs from that of the SSE,
at least neglecting purely clock-driven contribution ,that will be discussed in the next section.

As mentioned in Sec.~\ref{Early universe, ultralight scalar field and domain wall solution},
the inhomogeneous sign scenario implies that $\Delta \nu/\nu_0$ arises from a destructive sum of the fundamental variations  
$\Delta m_e/m_e$ and $\Delta \alpha/\alpha$.  
This results in a richer phenomenology for a given set of fundamental physical parameters  
associated with the DW and its interaction with ordinary matter.  
Indeed, the frequency variation can be highly suppressed  
if the single contributions in Eq.~\eqref{19} have nearly identical magnitudes,  
or can exhibit an opposite sign depending on the active atomic clock during the crossing event.
This consideration applies both to satellite clocks and to the reference clock and generates much more complicated domain wall
theoretical bias (or RFF) patterns.

The discussion presented so far has a purely theoretical  
and pedagogical scope. It is based on the assumption  
that it is possible to work with an almost continuous,  
smooth, and even constant GNSS clock bias.  
In practice, however, the GNSS clock bias is computed concurrently to a very sophisticated procedure 
known as Precise Orbit Determination (POD). 
As a consequence, the bias is a complicated function of the sampling variable $t$
and it exhibits two main features, 
\textit{namely} 
\textit{(i)} a sampling nature and  
\textit{(ii)} discontinuities and 
background noises affecting short, medium and long term behavior, \textit{i.e.} the stability of the space-born clocks.  
The following subsections are dedicated to an overview on the Precise Orbit Determination (POD) method useful to generate
clock bias time series, the clock bias times series properties
and the possibility to use the \emph{bias difference} time series to search for domain wall crossings.

\subsection{GNSS clock data: Precise orbit determination, clock bias generation and sampling}
\label{GNSS clock data: Precise orbit determination, clock bias generation and sampling}

The POD provides the reference recipe for reconstructing 
accurate satellite orbits, for review see \cite{Montenbruck:2000satellite}.
The procedure combines 
\textit{(i)} a set of tracking observations with
\textit{(ii)} a satellite model
and then estimates the orbit by minimizing the \emph{residuals} between the observed motion and the theoretical motion
predicted by the model. 
In this way, a best-fit trajectory is obtained together with the estimation of possible free parameters.
In the case of satellites equipped with atomic clocks, such as GNSS satellites,
the POD also provides clock bias solutions of the form Eq.~\eqref{21}.
Indeed, the clock bias is not obtained as a direct readout of the onboard clock. 
Rather, the GNSS signals received by ground stations carry timing information tied to 
the satellite clock at transmission, mixed with geometrical, propagation, instrumental, 
and relativistic effects, as well as station and satellite clock offsets with respect to 
the GNSS time scale. 
A POD procedure that accurately models these contributions can finally allow the construction 
of the satellite clock bias solution with respect to one (or more) ground station reference clock(s).

The accuracy of a POD solution (orbit and clock bias)
depends crucially on both the quality of the observations 
and the reliability of the satellite model. 
For GNSS satellites, the observations are mainly provided by microwave pseudorange and 
carrier-phase measurements collected by ground stations, possibly complemented by 
satellite laser ranging (SLR) data. 
The satellite model contains two main ingredients: a \emph{geometrical-structural} model and a 
\emph{dynamical} model. 
The former describes the spacecraft shape and its surface, thermal, and optical properties. 
Depending on the required accuracy, this description may range from a simple \emph{cannonball} 
approximation to a more realistic \emph{box-wing} (BW) model, or even to a \emph{finite-element} (FEM) model description 
that takes into account the fine properties for the spacecraft \cite{Lucchesi:2024zck,Sapio:2024pwu}.

The dynamical model describes the forces acting on the satellite. 
These are usually separated into gravitational and non-gravitational perturbations (NGP). 
The gravitational perturbations are largely spacecraft-independent and 
include the Earth's gravity field (typically up to order $(l_{\rm max}, m_{\rm max}) = (12,12)$ in the potential harmonic expansion, 
following the International Earth Rotation and Reference Systems Service (IERS) convention \cite{Petit:2010iers}),
tidal effects (solid and ocean tides)
associated with the response of the Earth to external bodies such as the Moon and the Sun, 
as well as relativistic corrections in the weak-field and slow-motion regime 
(Schwarzschild, de Sitter and Einstein-Lense-Thirring precessions, for a comprehensive review see \cite{Lucchesi:2026lrs}).
On the other side, the NGP depend on the structure of the satellite. 
They include the direct solar radiation pressure (SRP), Earth albedo and infrared radiation pressure (IRP),
thermal effects due to a plethora of phenomena of different origins.
The SRP provides the dominant acceleration source 
and its accurate modeling is essential to reduce orbit residuals and obtain
robust clock bias estimates \cite{Sapio:2024pwu}.

The complexity of POD requires the use of dedicated software packages
whose choice depends on the specific scientific goal. Among them, the BERNESE developed at the Astronomical Institute of the University of Bern (AIUB) \cite{dach2015bernese}, the GEODYN II \cite{mullins1972geodyn} developed by NASA/GSFC for GPS satellites and, for instance, NAPEOS \cite{dow2009napeos,springer2011napeos} used by the European Space Operations Center (ESOC), just to name a few. In particular, BERNESE and NAPEOS are multi-GNSS software and deal with microwave observations (pseudorange and carrier phase) and SLR data, whereas GEODYN II accepts a number of input data.
The POD software ensures that the resulting satellite clock bias estimates are sufficiently decoupled from orbital errors, providing a very good stability required in many physical measurement applications.
In the present work, the Galileo clock bias solutions obtained within the GREAT (Galileo gravitational
Redshift Experiment with eccentric sATellites) project 
via the NAPEOS software are used \cite{Delva:2018ilu,Herrmann:2018rva}. The details will be provided in the next section.

The feature of the POD procedures naturally provides a \textit{discrete} GNSS bias time series 
$S(t_k)$ characterized by a standard sampling window, typically $\tau = t_k - t_{k-1}$ of $30$ seconds. 
The presence of a such sampling 
could effectively convert a domain wall boxcar-like bias into a signal strongly dependent on (the connection between) 
the sampling epochs, the interaction times, and the domain wall transverse size. 
As a consequence, there is no guarantee to obtain a clear boxcar shape. 
For example, if the temporal 
extension of the boxcar is smaller than or of the order of the standard sampling window $\tau$, 
it becomes clearly impossible to accurately sample the entire plateau.
This signal could still be interpreted as a domain wall crossing or another nontrivial physical effect, but 
as discussed in the following subsections, one can instead work with the bias difference, \textit{i.e.} the (discrete) RFF.

\subsection{GNSS clock data: discontinuities and noise}
\label{GNSSS clock data: discontinuities and noise}

The GNSS bias of an atomic clock also exhibits two other notable features,  
\textit{i.e.} \textit{(i)} day-boundary discontinuities, 
jumps\footnote{Here, ``jump'' indicates an \emph{anomalous} large clock value away from day-boundaries.}
and data gaps,  
and \textit{(ii)} a noise-driven, nontrivial temporal evolution.

Day-boundary discontinuities are related to the specific features of their
estimation procedures \cite{RoviraGarcia:2021DBD} while jumps are usually connected to other station operations.
On the other side, data gaps may arise from a variety of reasons,
\textit{e.g.} limited tracking-network coverage or station, satellite and network-connection failures
\cite{Shi:2019ClockDiscontinuity}. 
In principle, data gaps could be addressed
using various methods, such as interpolation, extrapolation, or alternative
data sources, depending on the specific context.

The temporal evolution of the GNSS bias is also characterized  
by a stochastic and nonstationary process driven by a multi-component background clock noise 
that can easily mask the occurrence of an almost boxcar signal.
At short timescales (from sub-seconds to a few seconds) \cite{Hauschild:2013GNSSClock}, high-frequency fluctuations are mainly due to thermal noise,  
quantum noise, and short-term electronic instabilities.  
These induce random variations of the order of $\sim 10^{-10}$~s,  
whose mean value approaches zero when averaged over longer durations.  
On longer timescales (days to weeks) \cite{Huang:2013GPSClock}, low-frequency noise sources dominate.  
These include stochastic or random walk processes such as light-shift phenomena \cite{Formichella:2017AcStarkGPS} or slow thermal drifts, 
and aging effects associated with the gradual degradation of hardware components.  
The cumulative effect of these processes is a slow drift that can  
significantly affect the long-term stability of the atomic clock. 
It is also important to outline that GNSS bias of atomic clocks onboard satellites with elliptic orbits, such as E14 and E18 of the Galileo constellation,
also undergo a strong linear drift caused by the related astrodynamics in the Earth's gravitational field, which is practically negligible for satellites in standard nominal orbits.  
This drift dominates the evolution providing bias values of the order of $10^{-3}$~s over several years ($\geq 10^6$ samplings).

In general, the noise performance can be evaluated either in the frequency domain, 
via a power spectral density (PSD) analysis, or in the time domain 
through an Allan (variance or deviation) analysis, see \cite{Riley:2008FrequencyStability} for a review.
In particular, the Allan analysis 
\textit{(i)} provides a natural framework to highlight 
the contribution of different noise components across various timescales 
and 
\textit{(ii)} allows the identification of the bias stability, marking the transition 
from short- and medium-term fluctuations to long-term trends,  
which is particularly useful for conducting reliable and robust physical measurements.
This kind of analysis could be particularly effective  
for investigating \textit{persistent} physical processes over very long timescales.
However, a domain wall crossing is a purely local transient event 
that induces a collective effect across an entire network -- or, more in general, sensor network.
In this respect, it would be preferable to characterize a global \textit{clock or satellite noise network}, as discussed below.

\subsection{GNSS clock data: pseudofrequency and coincidence events}
\label{GNSS clock data: pseudofrequency and trigger event}

The previous discussion about the main properties of the bias time series 
raises some issues that need to be addressed, \textit{i.e.}, 
\textit{(i)} is it still reasonable to search for boxcar-like signals in the bias or one should consider another observable? 
\textit{(ii)} what is the impact of the sampling on this new variable and 
\textit{(iii)} what is the impact of the background noise?

\subsubsection{Pseudofrequency and domain wall signal amplitude}
\label{Pseudofrequency and domain wall signal amplitude}

The strongly sampled nature of the bias time series motivates the use of its discrete time derivative, 
namely the discrete RFF, commonly referred to as the \textit{GNSS bias difference} 
or \textit{GNSS pseudofrequency}:
\begin{equation}
S^{(1)}(t_k) = S(t_k) - S(t_{k-1}).
\label{34}
\end{equation}
This quantity converts large daily discontinuities into sharp and easily identifiable anomalies, 
removes possible linear trends in the data, 
and is expected to produce a pair of separated, non-normalized spikes with opposite signs 
when a large-size domain wall crosses the network, as also shown in the previous section.
\cite{Derevianko:2013oaa,Roberts:2017hla}\footnote{Note, however, that extremely large domain walls could also induce oscillations in the data, 
as concluded in \cite{Bertrand:2024ypu}.}.

Such pairs of bias difference spikes are expected to be highly localized in time,
in particular they should appear at a single sampling epoch.
This makes them intrinsically easier to detect than a broad, time-extended, boxcar-like signature 
in the original bias time series.

\begin{figure*}
    \centering
    \includegraphics[
      width=1\textwidth,
    ]{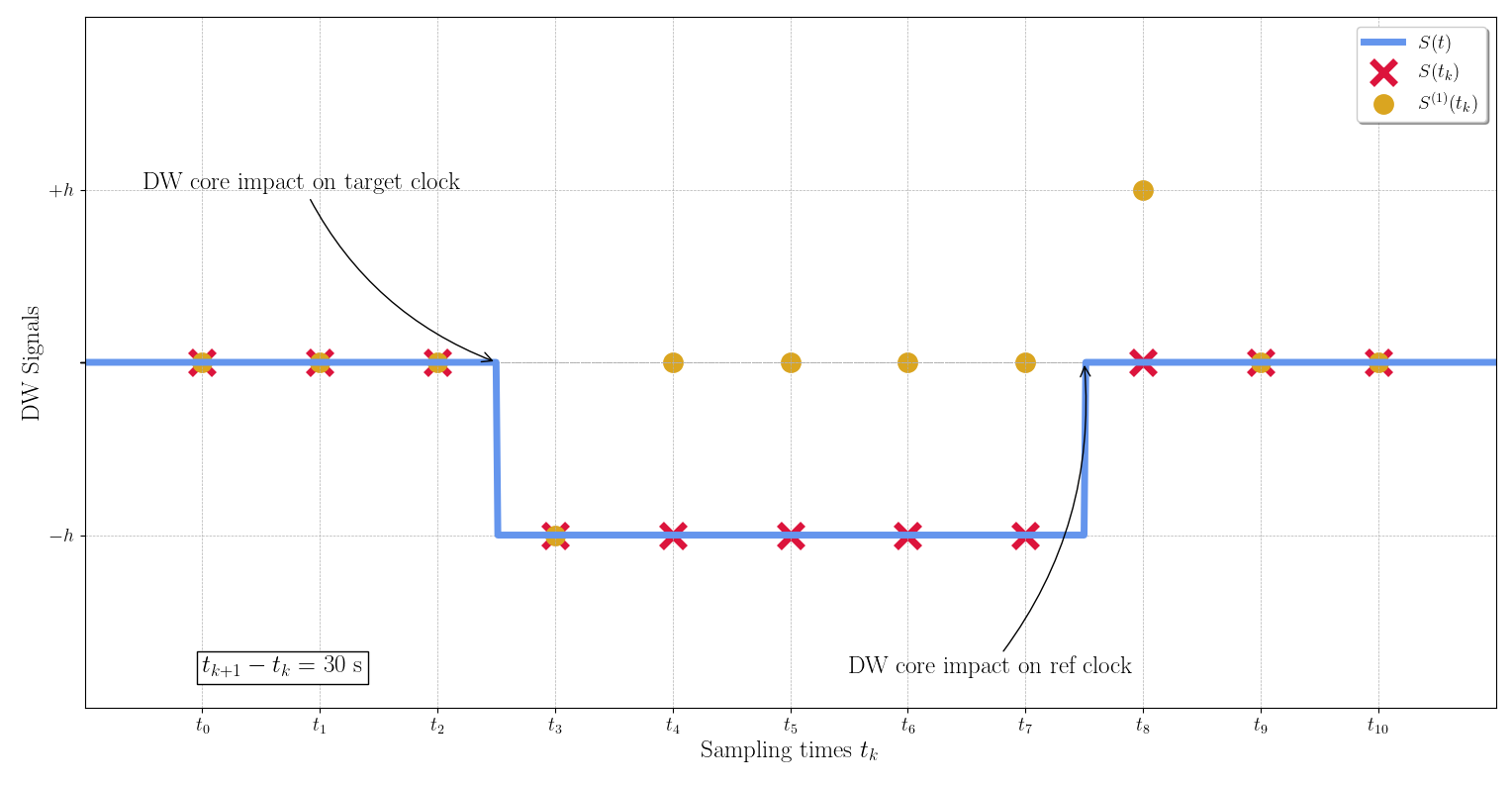}
    \caption{\it 
    Idealized effect of a DW crossing on the sampled clock bias $S(t_k)$ and on the corresponding discrete derivative (RFF)
    $S^{(1)}(t_k)$, with a standard sampling window of $t_{k+1}-t_k=30$ seconds.
    The RFF converts the two clock-interaction epochs into a pair of localized spikes.
    In the ideal case, the trigger and satellite spikes easily emerge from the background noise.
    }
    \label{Fig6}
\end{figure*}

Fig.~\ref{Fig6} shows a schematic example of the interaction between a domain wall and a satellite of a GNSS constellation.
A domain wall crossing can then be investigated by searching for plausible \emph{partial or complete coincidence events} 
in a set of bias difference time series occurring at a common reference time $t_{\text{ref}}$, 
as anticipated in Sec.~\ref{Basic theory}.

In that case, one might also consider a \emph{topological pattern} check, \textit{i.e.}
the hypothetical distribution of SSEs
at $t_{\text{sat},i}$, clustered around a given \textit{coincidence event},
 to obtain tighter bounds on the observation.
The expected \textit{average} amplitude of the single spikes (\textit{i.e.} SSE/GSSE) induced by the domain wall passage can be estimated from 
Eq.~\eqref{32} as \cite{Roberts:2017hla}
\begin{equation}
    S^{(1)}_{\text{avg}} \sim \frac{\rho_{\text{walls}}\, d^2\, \mathcal{T}}{\Lambda^2_e}\left(\frac{\Delta t}{\tau_{\text{avg}}}\right)
    \sim \frac{\rho_{\text{walls}}\, d^2\, \mathcal{T}}{\Lambda^2_e}
\label{35}
\end{equation}
and can also be referred to as a non-normalized version of Eq.~\eqref{32}.
Here, $\Delta t$ indicates the considered sampling window of the bias difference and the ratio $\Delta t/\tau_{\text{avg}}$
is a simple numerical factor $<10$ with velocities $v_g < 300$ km$/$s.

Consider a hypothetical scenario where the domain walls have size
$d \sim 10^3$ km, their local energy density represents a small part of the hypothetical dark matter density 
$\rho_{\text{walls}} \sim( 10^{-5}- 10^{-6})\rho^{\text{local}}_{\text{DM}} \mbox{ with }  \rho^{\text{local}}_{\text{DM}}  \sim 0.4$ GeV/cm$^3$,
induces a signal of the order of $10^{-11}$ seconds and the effective scalar-electron coupling is $\Lambda_e\sim 10^8$ GeV.
Then, the corresponding average encounter time is of the order of 
$\mathcal{T} \gtrsim 1$ year.

\subsubsection{Pseudofrequency and sampling}
\label{Pseudofrequency and sampling}

\begin{figure*}
    \centering
    \includegraphics[
      width=1\textwidth,
    ]{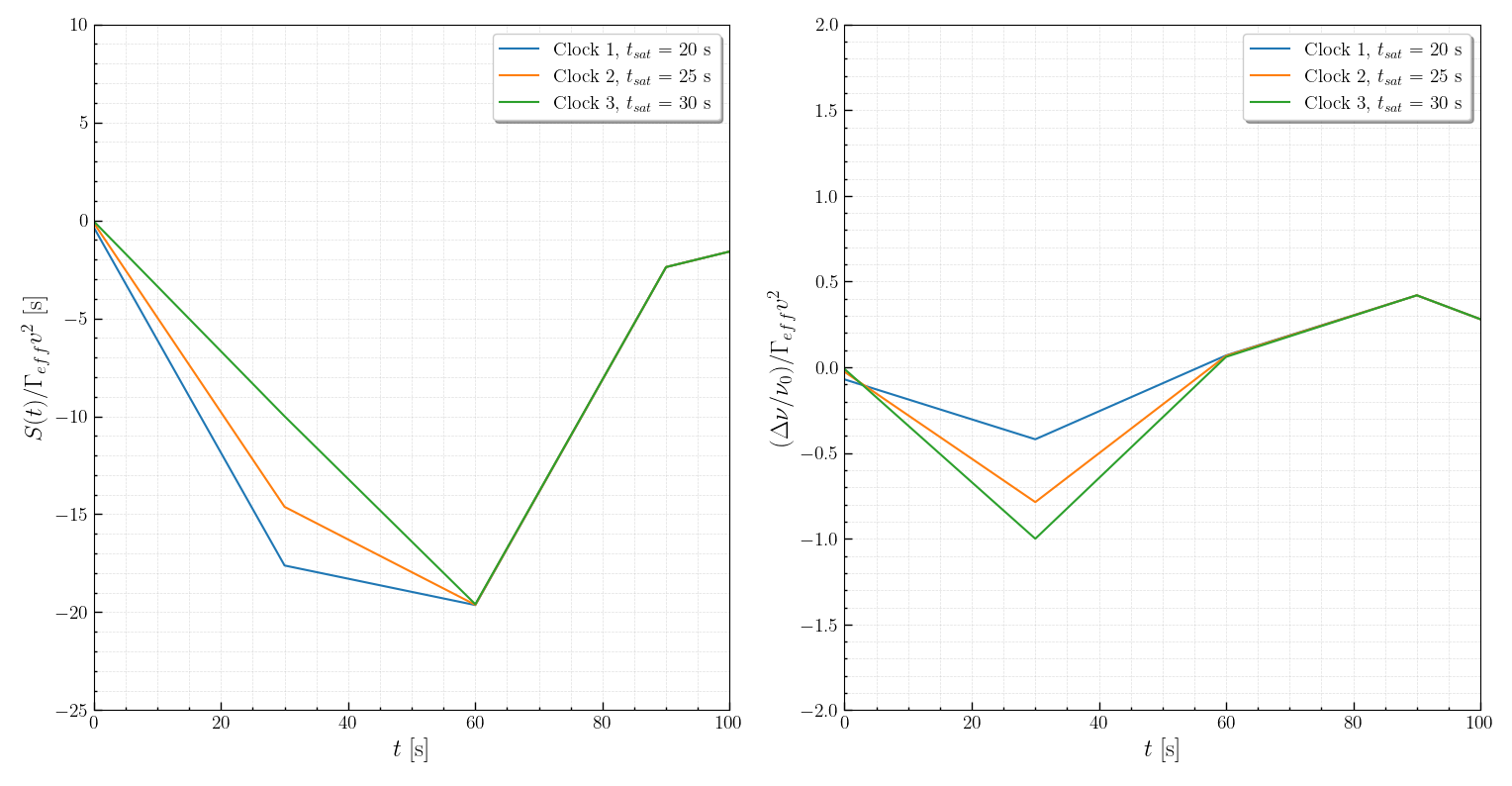}
    \caption{\it 
    Left panel: Sampled normalized clock-bias pattern for three satellite clocks interacting with the DW at different epochs.
    The sampling turns the smooth boxcar-like profiles into \emph{hat-like} reconstructed signals.
    Right panel: Corresponding normalized RFF, where the satellite-spike amplitudes depend on the relative distance
    between the sampling epochs and the DW-clock interaction times, while the trigger spike is the same.
    }
    \label{Fig7}
\end{figure*}

The sampling nature of the clock data affects the expected value of the bias difference in Eq.~\eqref{35}, \textit{namely}
the amplitude of both GSSE and SSE  
depends on the time difference between the GNSS sampling epoch  
and the interaction time with the domain wall.
In particular, the amplitude of the GSSE at $t_{\text{ref}}$  
is expected to be practically the same across all clock time series.  
On the contrary, this does not hold for the SSE, whose amplitude varies from satellite to satellite: 
the spike becomes sharper as the sampling epoch approaches the interaction time 
and may be smaller otherwise.
This suggests that the scenarios depicted in Fig.~\ref{Fig6}
and primarily discussed in \cite{Derevianko:2013oaa,Roberts:2017hla}
represent an ideal case that should hold under or during specific conditions.
Fig.~\ref{Fig7} shows the evolution of the bias and the corresponding bias difference in more realistic cases.
Unfortunately, the sampling process does not represent the main source of signal \textit{disturbance}.
As discussed below, this picture is severely modified by the presence of the intrinsic clock noise.

\subsubsection{Pseudofrequency and noise}
\label{Pseudofrequency and noise}

The background noise of an atomic clock network joins to the
sampling effects and significantly complicates matters.
In particular, it definitively represents the most important aspect to handle and can overwhelm,
partially overwhelm or even simulate a fundamental \textit{coincidence event}. 

In order to explore these scenarios, consider a network of $N$ space clocks  
and, for simplicity, just consider what happens to the hypothetical \textit{coincidence event} neglecting the sampling impact.

In the best case,  
the domain wall induces RFF variations  
that are large compared to the local noise fluctuations.  
A \emph{complete coincidence event} may then occur,  
with nearly identical GSSE amplitudes  
appearing across all clocks in the network.

In more realistic situations,  
some clocks may exhibit local noise fluctuations  
strong enough to interfere with the domain wall signal.  
This leads to only a \emph{partial coincidence event}, involving $M < N$ clocks.

In the worst case, noise dominates and prevents any potential coincidences.

Conversely, it is also possible 
that the network noise randomly generates
a pattern resembling a true \textit{coincidence event} (partial or complete).
Such accidental phenomena might be mistakenly interpreted  
as a signature of a domain wall passage.

These considerations lead to two essential questions:

\begin{enumerate}[label=\textit{\arabic*})]
    \item What is the probability that the network noise alone generates a \textit{partial or complete coincidence event}
    exceeding a given RFF threshold?

    \item What is the probability that a \textit{partial or complete coincidence event} remains detectable  
    despite the presence of a network noise?
\end{enumerate}

The next section will address these questions through a dedicated analysis  
of a selected Galileo GNSS satellite network.
As a final remark, the station operations mentioned 
in Sec.~\ref{GNSSS clock data: discontinuities and noise} could also have a relevant impact. 
However, their effects are typically clustered in time 
and should provide a pattern distinct from the domain wall one.
In particular, they tend to involve all the operating satellites
producing complete coincidences but with a related distribution of the satellite spikes
incompatible with a propagating phenomena across the constellation.
Therefore, this kind of events could be recognized and discarded during a concrete domain wall search
by using topological pattern checks, as mentioned in \ref{Pseudofrequency and domain wall signal amplitude}.

\begin{figure*}
    \centering
    \includegraphics[
      width=1\textwidth,
    ]{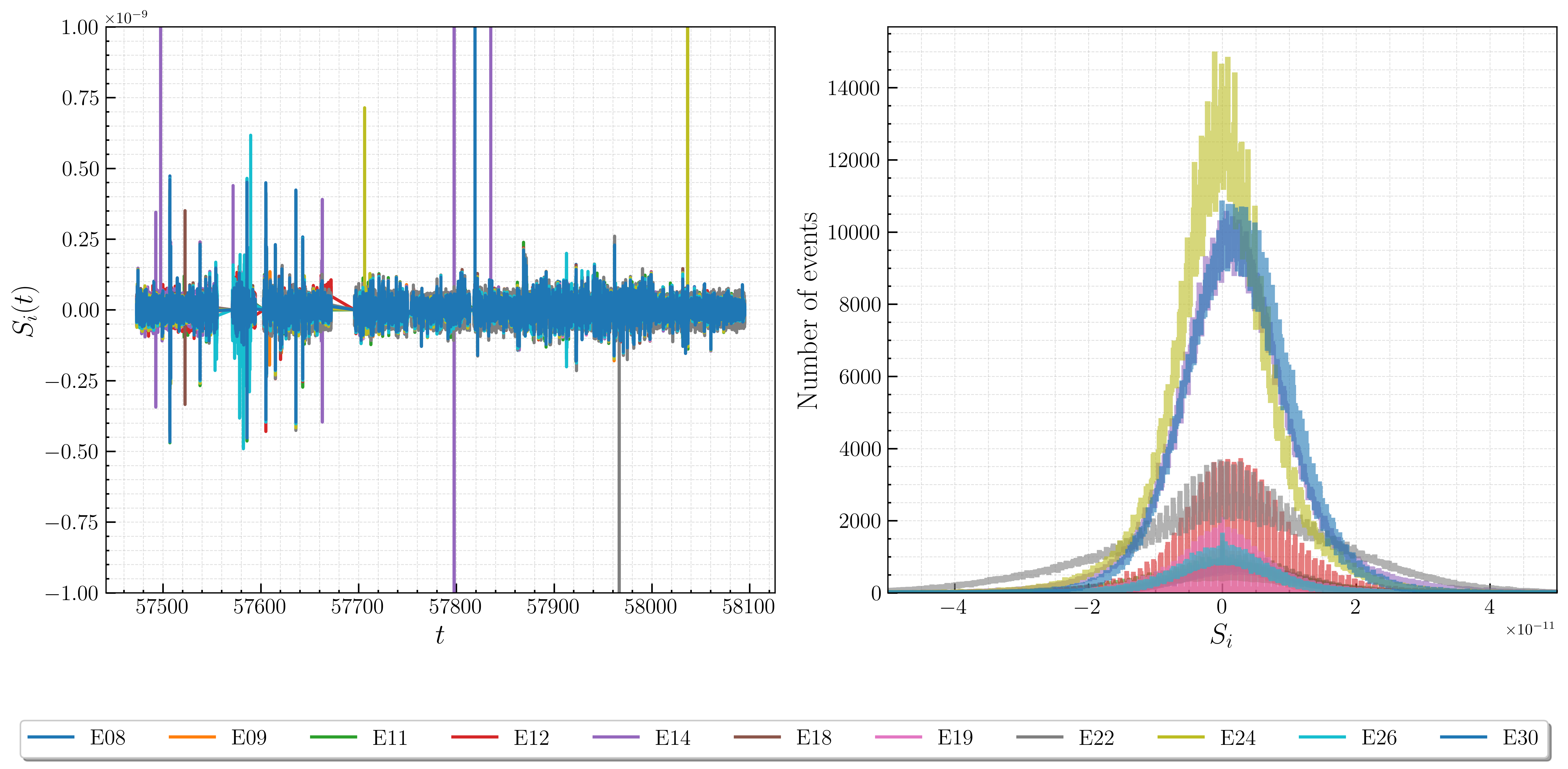}
    \caption{\it Left panel: time series of the bias differences for the selected eleven Galileo satellites over the common analysis period, after the cleaning procedure. The independent variable \textit{t} is in this case in the Modified Julian Date (MJD). The time series exhibit strong oscillations around the expected value
    with particularly noisy periods. 
    Data gaps are due to the discontinuous coverage of the overall observation period. Local spikes with amplitudes larger than a nanosecond are also present, likely caused by various effects such as ground-station operations. The type of onboard master clock used by each satellite, Rb or PHM, over a given period also affects the local properties of the time series. 
    Right panel: histograms of the bias-difference time series. 
    The bias-difference values are generally distributed around zero, although with satellite-dependent distributions.}
    \label{Fig8}
\end{figure*}

\begin{figure*}
    \centering
    \includegraphics[
      width=1\textwidth,
    ]{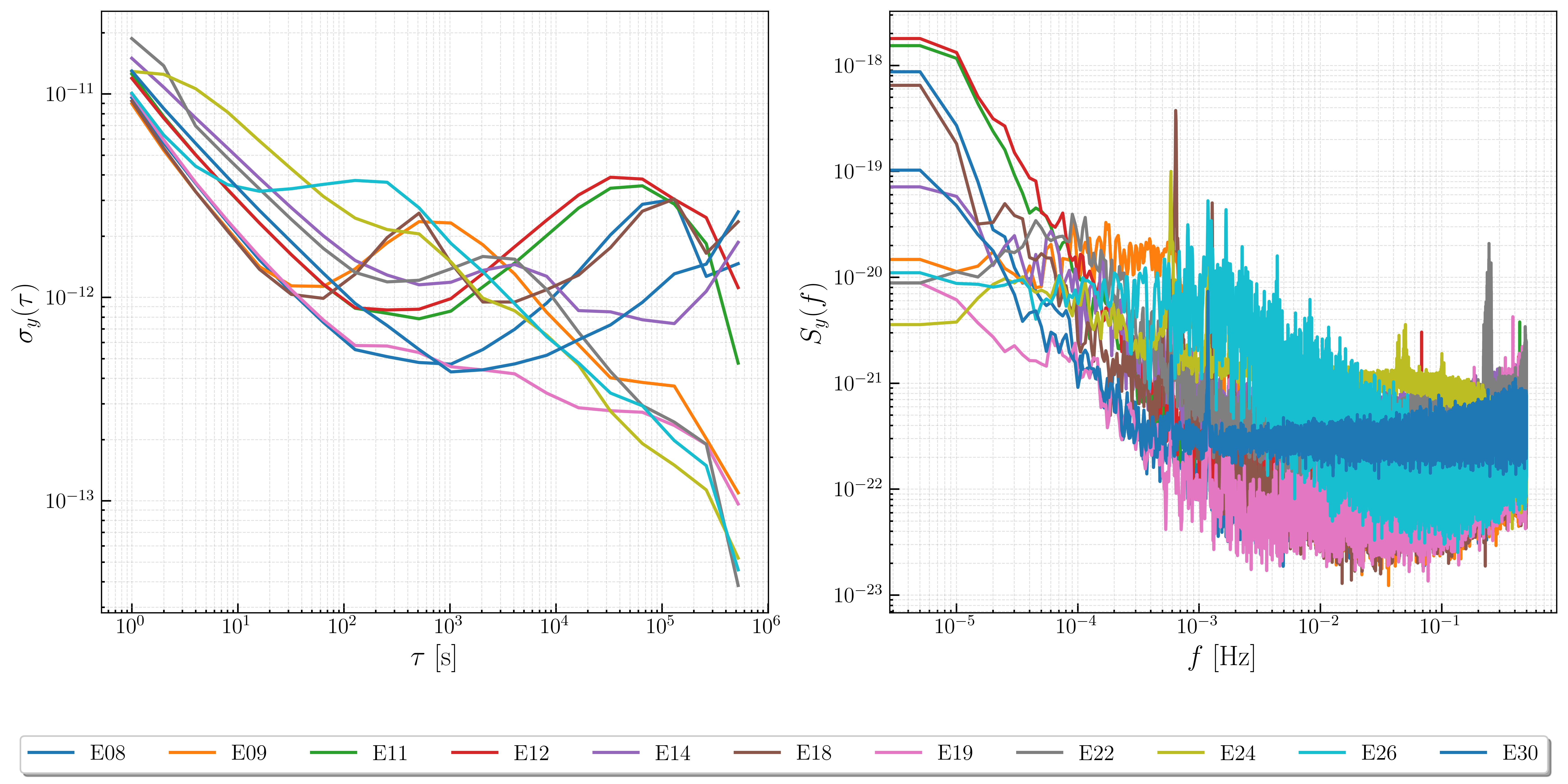}
    \caption{\it Left panel: overlapping Allan deviation analysis for the eleven time series of the normalized bias difference. The resulting Allan deviation curves differ from satellite to satellite and are affected by the onboard clock switching
    and GNSS operations. Right panel: power spectral density analysis of the normalized time series. 
    The PSD curves are globally consistent with the time stability results 
    and outline the distinctive features of the RFF time series.}
    \label{Fig9}
\end{figure*}

\section{Domain walls and Galileo GNSS atomic clock noise}
\label{Domain walls and Galileo GNSS atomic clock noise}

The answers to the questions just raised in the previous section require two complementary analyses: a false-alarm rate (FAR) analysis, 
aimed at estimating the probability that background noise alone may generate plausible \textit{partial or complete coincidence events} 
and a detection-efficiency analysis, designed to quantify the response of the clock network to a domain wall crossing 
and the corresponding probability of producing \textit{partial and complete coincidences}.

These analyses were performed on a set of eleven Galileo satellites, 
considering only the periods during which all of them were operating simultaneously, 
within the 2014--2017 three-year dataset, specifically from 26 March 2016 to 7 December 2017, for a total of 578 days.
The bias difference time series were first subjected to a cleaning procedure 
in order to remove residual linear trends and explicit daily discontinuities.
Fig.~\ref{Fig8} reports the resulting time series and their related distributions
whereas Fig.~\ref{Fig9} the associated stability analysis in the time and frequency domains.
These figures show the very complicated behavior of the bias (difference) time series
-- with some periods less noisy than others --
the emergence of auto and cross correlations or even clock switching.

The analyses were implemented under the assumption that the various
(Earth-sized walls-induced) GSSEs, at the same $t_{\text{ref}}$, 
are not expected to have exactly the same amplitude,
because of the sampling procedure and the local noise properties, previously mentioned. 
As discussed below, the individual GSSEs were simply required to reach anomalously large values, 
exceeding suitably defined threshold levels, as done in \cite{Roberts:2017hla} for the GPS constellation.

\subsection{False alarm rate analysis}

The goal of the false-alarm analysis is to quantify the probability that background noise alone produces 
simultaneous crossings in $n = 2,3,\dots,N$ satellites for a set of chosen thresholds $\{\theta_i\}$ over the 
selected observed period.
The thresholds determine a binary result for each satellite at each epoch, \textit{i.e.} signal-present or signal-absent state, 
while the collection of the occurring GSSEs, \textit{i.e.} the multiple crossing events,
are referred to as a \emph{coincidence} and range from the \emph{2-fold} coincidence (or 
\emph{2-fold} partial coincidence) 
up to the maximal \emph{$N$-fold}, complete and spurious coincidence.

In this context, the FAR analysis was implemented via two complementary approaches.
The first is a theoretical prediction based on the individual tail probabilities of each satellite (to be above each $\theta_i$) and on
a set of statistical assumptions. 
The second is an empirical determination obtained by a \emph{time-slide} procedure
applied to the original time-series dataset. 
This procedure also allows testing the theoretical recipe when probed against the experimental data.

It is important to stress that, as mentioned in Sec.~\ref{Basic theory}, 
each Galileo satellite is equipped with (a pair of) two onboard clocks,
-- a PHM and a RAFS -- and a satellite bias time series is a sequence of different clock contributions.
However, in the present analysis each satellite is treated as a single node of the GNSS network
-- without explicitly accounting for which onboard clock is the operational one at a given time -- 
since PHM and RAFS clocks are assumed to share the same basic statistical properties.

The theoretical framework relies on several working hypotheses.
In particular, the background noise is assumed to be statistically independent from satellite to satellite, 
implying negligible cross-correlations between different network nodes.
The noise associated with each individual satellite is also assumed to exhibit only weak temporal autocorrelation, 
so as to prevent artificial clustering of spurious spikes on specific (eventually short) timescales.
In addition, the noise statistics are assumed to be almost stationary within the selected period.

From a mathematical point of view,
each satellite node $i=1,\dots,N$ is modeled via a Bernoulli random variable
\begin{equation}
X_i(t_k, \theta) =
\begin{cases}
1 & \text{if } A_i > \theta, \\
0 & \text{if } A_i \le \theta,
\end{cases}
\label{36}
\end{equation}
where $A_i$ denotes the measured bias difference amplitude at a given epoch $t_k$. 
Therefore, 
this variable describes if the satellite bias difference value crosses a selected threshold $\theta$
taken from the chosen set $\{\theta_i\}$.
If the events are rare, the corresponding tail probability for the satellite $i$ to be found above a $\theta$ value at some sampling epoch $t_k$
is denoted by
\begin{equation}
p_i(t_k, \theta) = \mathbb{P}(A_i > \theta),
\label{37}
\end{equation}
and is estimated as
\begin{equation}
p_i(t_k, \theta) =
\frac{\text{Number of times }  A_i > \theta}{N_{\mathrm{rows}}}.
\label{38}
\end{equation}
where $N_\text{rows}$ is the number of sampling epochs, \emph{i.e.} the length of the observed period.
A complete tail probability curve can be computed by considering the entire set of chosen threshold values $\{ \theta_i \}$.

Then, one can introduce the collective network variable
\begin{equation}
S_N(t_k, \theta) = \sum_{i=1}^{N} X_i(t_k, \theta),
\label{39}
\end{equation}
representing the total number of satellites exceeding $\theta$ at an epoch $t_k$. 
The probability of observing exactly $n$ satellites above $\theta$ for $t=t_k$ can then be written as
\begin{equation}
P_N(n \mid t_k, \theta) = \mathbb{P}\!\left(S_N(t_k, \theta)=n\right).
\label{40}
\end{equation}

The assumed statistical independence among the $X_i$
allows one to compute the theoretical realization of Eq. ~\eqref{40}, 
\textit{i.e.} the probability mass function $P^{\text{th}}_N(n \mid t_k, \theta) = \mathcal{P}_n(\theta)$
as a Poisson-Binomial distribution \cite{Biscarri:2018}
\begin{equation}
\mathcal{P}_n(\theta) = 
\sum_{\substack{U \subset \{1,\dots,N\} \\ |U| = n}}
\left(
\prod_{i \in U} p_i(\theta)
\right)
\left(
\prod_{j \notin U} \bigl(1 - p_j(\theta)\bigr)
\right),
\label{41}
\end{equation}
where the sum extends over all subsets $U$ of cardinality $n$.
This expression provides the theoretical false-alarm probability 
of a rare $n$-fold coincidence event over the observation time span
under the noise-only hypothesis and can be quantitatively estimated by convolution recipe \cite{Biscarri:2018}.
In the special case of statistically identical (satellite) tail probabilities 
--
$p_i(\theta)=p(\theta),\, \forall i$
--
the distribution simplifies to the binomial one
\begin{equation}
\mathcal{P}_n(\theta)
=
\binom{N}{n}
p(\theta)^n
\left(1 - p(\theta)\right)^{N-n}.
\label{42}
\end{equation}
In the limit of an extremely large network, \textit{i.e.} large $N$, and small single-node probabilities, $p(\theta_i)\ll1$,
the mass functions asymptotically approaches the pure Poisson statistics.
However, this approximation is not well suited in the present case, 
since the number of available (Galileo GNSS) satellites is intrinsically limited.

The empirical FAR was estimated directly from the time-series dataset.
In particular, a \emph{time-slide} procedure was used to generate a large number 
$N_{\rm shift}$ of randomly, locally, and circularly shifted realizations of the original dataset.
This local approach suppresses physical coincidences while, ideally, preserving 
the (more realistic) \emph{locally stationary} properties of the background noise.
Conversely, arbitrarily large shifts -- although compatible with the sample size --
would mix different epochs of the dataset,
potentially characterized by different noise properties, leading to a time-averaged FAR estimate.

The empirical probability $P^{\rm emp}_N(n \mid t_k, \theta) = P_n (\theta)$
that at some epoch $t_k$ of the realization $r$,
a $n$-fold \textit{coincidence event} occurs above a threshold $\theta$
is defined as
\begin{equation}
    P_{n}^{(r)}(\theta)
    =
    \frac{1}{N_{\rm rows}}
    \sum_{k=1}^{N_{\rm rows}}
    \mathbf{1}\!\left[
    S_N^{(r)}(t_k;\theta)=n
    \right],
\label{43}
\end{equation}

The corresponding mean value over $N_{\rm shift}$ realizations is
\begin{equation}
    \overline{P}_{n}(\theta)
    =
    \frac{1}{N_{\rm shift}}
    \sum_{r=1}^{N_{\rm shift}}
    P_{n}^{(r)}(\theta)
\label{44}
\end{equation}  
while the standard deviation over the $N_{\rm shift}$ realizations and the standard error on the mean are
\begin{equation}
    \sigma_{n}(\theta)
    =
    \left[
    \frac{1}{N_{\rm shift}-1}
    \sum_{r=1}^{N_{\rm shift}}
    \left(
    P_{n}^{(r)}(\theta)
    -
    \overline{P}_{n}(\theta)
    \right)^2
    \right]^{1/2}
\label{45}
\end{equation}
and 
\begin{equation}
    \sigma_{\overline{P}_{n}}(\theta)
    =
    \frac{\sigma_{n}(\theta)}
    {\sqrt{N_{\rm shift}}}.
\label{46}
\end{equation}
respectively. 
Note that the empirical false-alarm curves are bounded by the size of the dataset,
$N_{\rm rows}$, and by the number of time-slide realizations performed, $N_{\rm shift}$.
Accordingly, the smallest non-zero probability that can be empirically resolved
scales as $\sim (N_{\rm rows}N_{\rm shift})^{-1}$.

\begin{figure*}
    \centering
    \includegraphics[
      width=1\textwidth,
    ]{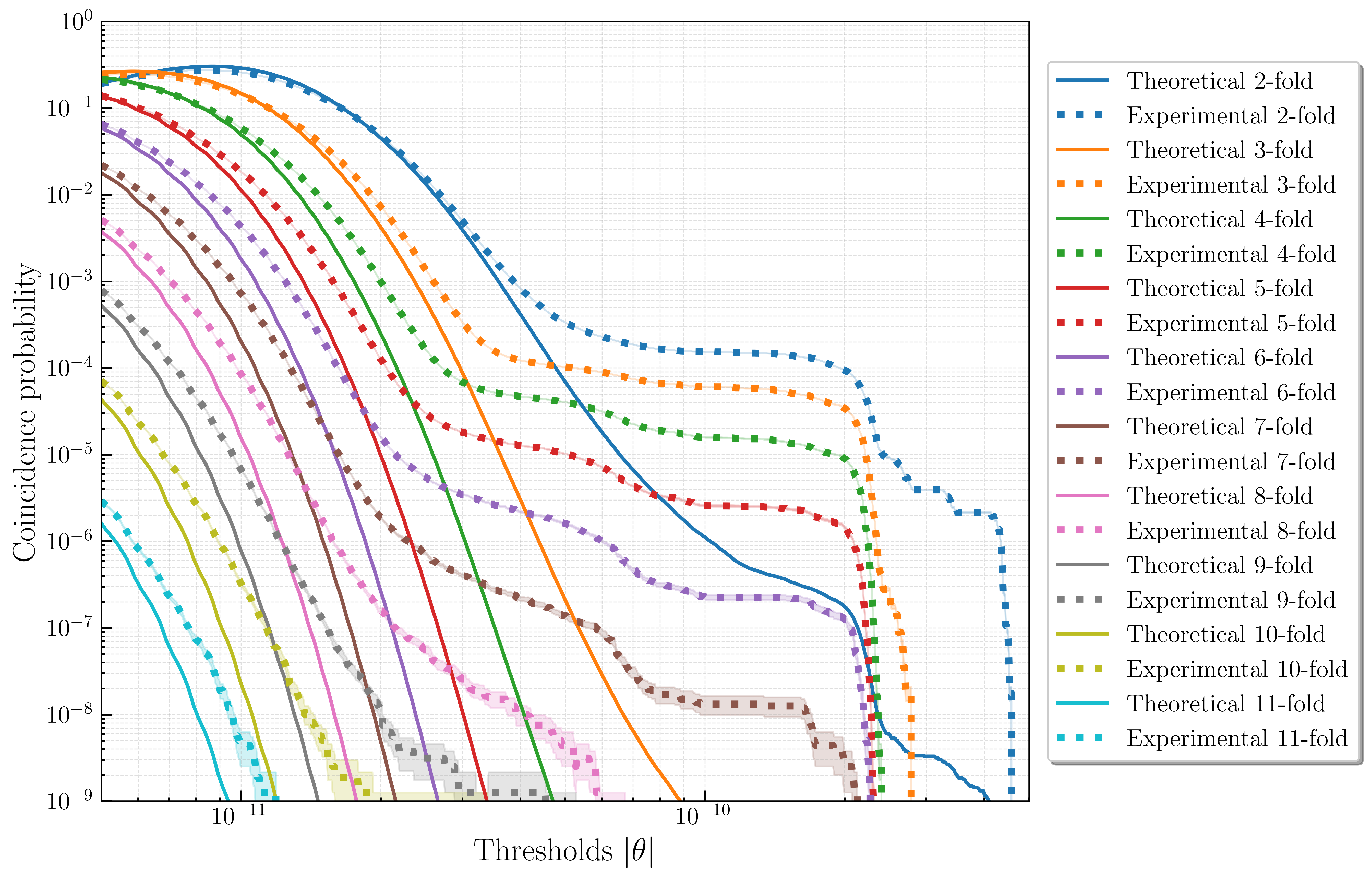}
    \caption{\it Experimental and theoretical false-alarm analysis conducted for positive spurious spike signals over the entire period.
    The experimental curves were computed using $N_{\rm cross}=10^3$ local circular shifts
    spanning over $10^3$ thresholds. }
    \label{Fig10}
\end{figure*}

Empirical false-alarm curves with uncertainty bands $\overline{P}_{n}(\theta) \pm \sigma_{\overline{P}_{n}}(\theta)$
can be generated and compared with the theoretical FAR, 
providing a consistency test of the theoretical statistical assumptions.
Fig.~\ref{Fig10} shows the results of the comparison conducted for \textit{(partial or complete) coincidence events}. 
If a negative threshold is assumed, similar outcomes are provided.
The results generically show systematic deviations
of the theoretical approach from the empirical one
confirming the presence of unmodeled correlations and therefore departures from the adopted independence hypothesis.
In particular, low-order coincidences ($\lesssim 5$) exhibit only mild
departures at sufficiently low thresholds ($\lesssim 10^{-11}$), whereas the
discrepancy gets more pronounced for higher coincidence orders.

The (empirical) FAR analysis shown in Fig.~\ref{Fig10} allows to identify the threshold
conditions under which the probability of noise-driven events is sufficiently
low to suppress the expected background.
For example, one can observe that the probability 
that the background noise could lead to a \textit{complete coincidence events, -- i.e. the maximal coincidence order (the eleven fold) --} 
above $10^{-11}$ seconds is of the order of $10^{-8}$
whereas the corresponding probability for a ten fold is just below $10^{-6}$.
Table~\ref{Tab2} reports threshold values corresponding to probability levels order of $10^{-7}$ and $10^{-8}$ 
for coincidences larger than the $n=6$-fold and for both type of signal signs.
In principle, one can expect that once a probability level is chosen, the corresponding threshold outcomes
are very close for positive and negative scenarios.
The results of Table~\ref{Tab2} shows no exact symmetry, suggesting a different statistical behavior for the background clock noise tails.

\begin{table*}[t]
\centering
\small
\renewcommand{\arraystretch}{1.25}

\begin{tabular*}{\textwidth}{@{\extracolsep{\fill}} l c c @{}}
\hline\hline

\parbox[t]{0.18\textwidth}{N-folds}
& \parbox[t]{0.34\textwidth}{Thresholds -- Positive triggers [s]}
& \parbox[t]{0.34\textwidth}{Thresholds -- Negative triggers [s]} \\
\hline

\multicolumn{3}{@{}l}{\textbf{Mean false-alarm probability level: \(10^{-7}\)}} \\
\hline

\parbox[t]{0.18\textwidth}{7-fold}
& \parbox[t]{0.34\textwidth}{\(5.8619 \times 10^{-11}\)}
& \parbox[t]{0.34\textwidth}{\(-6.5779 \times 10^{-11}\)} \\

\parbox[t]{0.18\textwidth}{8-fold}
& \parbox[t]{0.34\textwidth}{\(2.1558 \times 10^{-11}\)}
& \parbox[t]{0.34\textwidth}{\(-2.4869 \times 10^{-11}\)} \\

\parbox[t]{0.18\textwidth}{9-fold}
& \parbox[t]{0.34\textwidth}{\(1.5327 \times 10^{-11}\)}
& \parbox[t]{0.34\textwidth}{\(-1.2456 \times 10^{-11}\)} \\

\parbox[t]{0.18\textwidth}{10-fold}
& \parbox[t]{0.34\textwidth}{\(1.1359 \times 10^{-11}\)}
& \parbox[t]{0.34\textwidth}{\(-8.5350 \times 10^{-12}\)} \\

\parbox[t]{0.18\textwidth}{11-fold}
& \parbox[t]{0.34\textwidth}{\(7.7475 \times 10^{-12}\)}
& \parbox[t]{0.34\textwidth}{\(-5.4326 \times 10^{-12}\)} \\

\hline

\multicolumn{3}{@{}l}{\textbf{Mean false-alarm probability level: \(10^{-8}\)}} \\
\hline

\parbox[t]{0.18\textwidth}{7-fold}
& \parbox[t]{0.34\textwidth}{\(1.6615 \times 10^{-10}\)}
& \parbox[t]{0.34\textwidth}{\(-8.8760 \times 10^{-11}\)} \\

\parbox[t]{0.18\textwidth}{8-fold}
& \parbox[t]{0.34\textwidth}{\(4.1485 \times 10^{-11}\)}
& \parbox[t]{0.34\textwidth}{\(-5.1998 \times 10^{-11}\)} \\

\parbox[t]{0.18\textwidth}{9-fold}
& \parbox[t]{0.34\textwidth}{\(2.0210 \times 10^{-11}\)}
& \parbox[t]{0.34\textwidth}{\(-2.1757 \times 10^{-11}\)} \\

\parbox[t]{0.18\textwidth}{10-fold}
& \parbox[t]{0.34\textwidth}{\(1.4042 \times 10^{-11}\)}
& \parbox[t]{0.34\textwidth}{\(-1.1306 \times 10^{-11}\)} \\

\parbox[t]{0.18\textwidth}{11-fold}
& \parbox[t]{0.34\textwidth}{\(9.6218 \times 10^{-12}\)}
& \parbox[t]{0.34\textwidth}{\(-7.7833 \times 10^{-12}\)} \\

\hline\hline
\end{tabular*}

\vspace{0.15cm}

\parbox{\textwidth}{
\caption{\it Threshold values, in seconds, associated with two selected levels of the mean experimental false-alarm probability,
\(10^{-7}\) and \(10^{-8}\), for different coincidence orders. 
The second and third columns refer to spurious positive and negative triggers, respectively. 
As expected, moving to a more stringent false-alarm probability level requires larger threshold values for a fixed coincidence order. 
The absence of an exact symmetry between the positive- and negative-trigger thresholds reflects the different statistical behavior of the two tails of the clock-noise distribution.
}
\label{Tab2}
}

\end{table*}

\subsection{Detection efficiency analysis}

The detection efficiency study for the eleven-satellites Galileo network
was carried out by simulating multiple DW crossings.
The simulation setup takes into account
\textit{(i)} the keplerian parameters of the satellites and their rates, in the Earth-Centred Inertial (ECI) reference system, 
from the ESA Galileo Metadata\footnote{\url{https://www.gsc-europa.eu/system-service-status/orbital-and-technical-parameters}},
\textit{(ii)} the ground reference clock positions and
associated operating time periods,
\textit{(iii}) a set of DW model parameters (\textit{i.e}, signal amplitudes, DW velocity and incoming direction) and 
\textit{(iv)} the clock bias (difference) time-series referred to the common working periods of the eleven satellites.

The corresponding large number of synthetic events generated -- having different amplitudes --
was then added to the physical background and a dedicated selection recipe
finally allow to construct 
the probability of detecting \textit{at least} $n$-fold \textit{coincidences} above selected thresholds, 
as function of the threshold amplitude itself.

In general, a significant number of clocks are expected to cross a threshold 
simultaneously when the search threshold becomes comparable to, or smaller than, the absolute amplitude of the induced signal.
In this regime, the network can therefore reveal the presence of the injected signal through a \textit{partial coincidence event}.

The situation is different for a \textit{complete coincidence}, in which all satellites of the network cross the threshold at the same sampling epoch. 
This represents the most stringent coincidence condition but it is also statistically less likely, 
since it can be affected by several effects, ranging from local noise features to correlations among the impact times.
Indeed, in a large network, the probability that the domain wall crosses one or more satellites and the selected ground reference clock at nearly the same time (or within the same sampling window) increases.
In such cases, the corresponding positive and negative spikes may overlap, 
making it difficult to resolve them as two distinct signals, as anticipated in Sec. \ref{Domain wall dark matter and GNSS atomic clocks}.

\begin{figure*}
    \centering
    \includegraphics[
      width=1\textwidth,
    ]{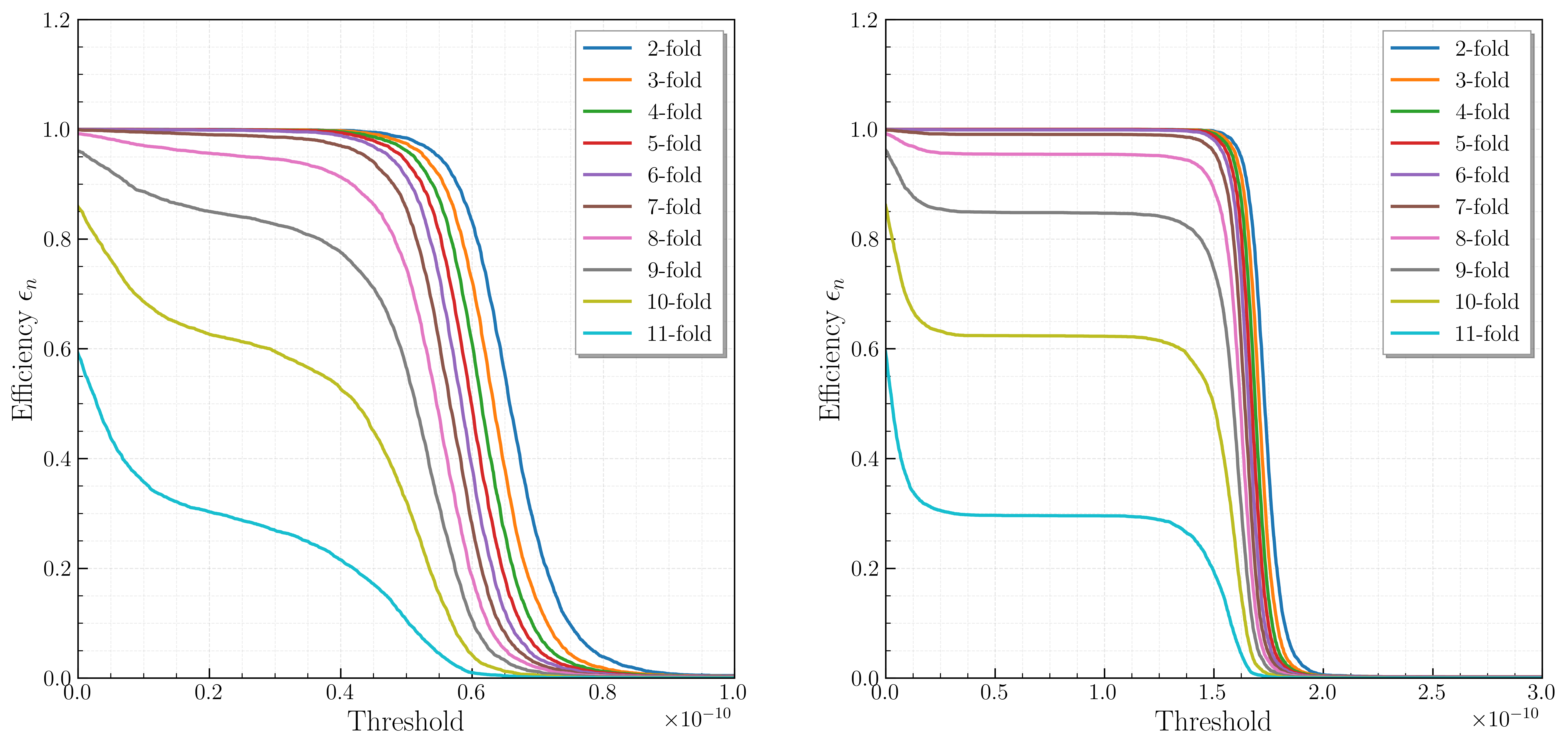}
    \caption{\it Left panel: detection efficiency analysis for a positive injected amplitude $5.86 \times 10^{-11}$ seconds.
    Right panel detection efficiency analysis for a positive injected amplitude $1.66 \times 10^{-10}$ seconds. 
    The two simulations show a very similar behavior for sufficiently low thresholds
    and then deviates as one explores larger values, typically due to the background noise contributions.}
    
    \label{Fig11}
\end{figure*}

Fig.~\ref{Fig11} reports the detection efficiency curves $\epsilon_{n}(A)$ 
for two injected amplitudes $A$ and for $N_{\rm cross}\sim 10^3$ 
simulated domain wall crossings through the Galileo GNSS geospace, with $v_g\sim 270$ km$/$s.
As mentioned, the $n$-fold curve represents
the evolution of the probability for \emph{at least} $n$ (satellite) crossings at the same epoch, given a selected threshold.
The left panel shows the results for positive GSSEs
with amplitude $A$ compatible with a threshold associated to a $7$-fold at a probability level $10^{-7}$ ($A = 5.86 \times 10^{-11}$)
while the right panel an amplitude compatible with the threshold associated to the a $7$-fold at $10^{-8}$ ($A \sim 1.66 \times 10^{-11}$), see Table~\ref{Tab2}.
In general, the network response shows a clear threshold-driven evolution. 
Particularly, at thresholds above the injected amplitude, the efficiency is strongly suppressed
and depends only on rare spurious noise-driven events.
Lowering the threshold towards the signal amplitude or just below, leads to a rapid increase of the efficiency, 
especially for the smaller low-order coincidences.

In the left panel case, the
$7$-fold and $8$-fold coincidences show a detection probability $<0.4$ 
for a searching threshold close to the injected signal.
In particular, the efficiency becomes larger than $0.9$ only going to very small searching-thresholds ($\lesssim 0.2\times 10^{-10}$ seconds).
The results for coincidence $>9$ are even more dramatic.
For searching-thresholds value $\sim 5\times 10^{-11}$ the efficiency for the $11$-fold is order $0.1$, for instance.

In the right panel case, the $7$-fold and $8$-fold coincidences become relatively good benchmarks of a domain wall crossing,
assuming a searching-threshold of order $\sim 1.5 \times 10^{-10}$ seconds.
Coincidences higher than $>9$-fold turn out to be more problematic.
The $10$-fold and $11$-fold efficiencies approach a constant efficiency value ($\gtrsim 0.6$ and $\sim 0.3$, respectively) 
significantly below the unity and
as one explores extremely small thresholds, they show a rising trend due to purely noise-driven contributions.
This effect is less marked in the left plot.

These two examples show that the detection of a \textit{complete coincidence} in the network is limited 
not only by the signal amplitude but also by the detailed features of the crossing kinematics, as mentioned above.

\subsection{Towards better constraints of domain wall crossings}
\label{Towards a better constrain of domain wall crossings}
The false-alarm and detection-efficiency analyses would suggest that the $n=7$-fold coincidence, 
with a nominal searching-threshold of order
$\theta\simeq 1.5\times 10^{-10}$ seconds, associated with an injected
signal amplitude $A=1.66\times 10^{-10}$ seconds, provides a favorable
channel and a compromise between background suppression and event recovery.
Under the
assumptions discussed in Secs.~\ref{Domain wall dark matter and GNSS atomic clocks},
this $(n,\theta)$-channel would therefore represent a natural starting point for a
domain wall search in the Galileo clock-bias data.

In particular, one may ask what kind of constraints could be inferred, at least as an \emph{illustrative exercise}, 
by assuming a Poisson statistics for rare events, as indicated by \emph{Roberts et. al.} in ~\cite{Roberts:2017hla}.
In this description, the probability
of observing $N_{\rm obs}$ domain wall-compatible signals in the given search
channel is modeled as
\begin{equation}
    P(N_{\rm obs}\mid n,\theta,A)
    =
    \frac{\mu_{n,\theta}^{N_{\rm obs}}}{N_{\rm obs}!}
    e^{-\mu_{n,\theta}} ,
\end{equation}
where $\mu_{n,\theta}$ denotes the total expected number of events in the
selected channel. This can be written as the sum of a background contribution $b_{n,\theta}$
and a physical signal contribution $s_{n,\theta}(A)$,
\begin{eqnarray}
    \mu_{n,\theta} &=& b_{n,\theta} + s_{n,\theta}(A), \nonumber \\
    b_{n,\theta} &=& \overline{P}_{n}(\theta) N_{\rm rows}, \nonumber \\
    s_{n,\theta}(A) &=& R\,T_{\rm obs}\,\epsilon_{n,\theta}(A).
\end{eqnarray}
Here $\overline{P}_{n}(\theta)$ is the false-alarm probability per sampling epoch,
$N_{\rm rows}$ is the number of sampled epochs, $R$ is the domain-wall
encounter rate, $T_{\rm obs}$ is the observation time, and
$\epsilon_{n,\theta}(A)$ is the detection efficiency obtained with an injected amplitude $A$.
The mean encounter time is defined as
\begin{equation}
    \mathcal{T} = \frac{1}{R}.
\end{equation}

If the selected channel could be treated as effectively background-free,
$b_{n,\theta}\ll 1$, and if no event were found after a reliable search supported by topological pattern checks,
$N_{\rm obs}=0$, one would get
\begin{equation}
    P(0\mid n,\theta,A)
    =
    \exp\left[-R\,T_{\rm obs}\,\epsilon_{n,\theta}(A)\right].
\end{equation}
An upper limit on the encounter rate at confidence level $CL$ would then be
obtained imposing
\begin{equation}
    \exp\left[-R_{CL}\,T_{\rm obs}\,\epsilon_{n,\theta}(A)\right]
    =
    1-CL,
\end{equation}
namely
\begin{equation}
    R_{CL}(A)
    =
    \frac{-\ln(1-CL)}
    {T_{\rm obs}\,\epsilon_{n,\theta}(A)} .
\end{equation}
Equivalently, this would imply a lower bound on the mean encounter time,
\begin{equation}
    \mathcal{T}_{CL}(A)
    =
    \frac{T_{\rm obs}\,\epsilon_{n,\theta}(A)}
    {-\ln(1-CL)} .
\end{equation}
By assuming a $95\%$ confidence level and a detection efficiency $\sim 1$ (for the used searching-threshold), 
this would correspond to
\begin{equation}
    R < R_{95} \simeq \frac{3}{T_{\rm obs}},
    \qquad
    \mathcal{T} > \mathcal{T}_{95} \simeq \frac{T_{\rm obs}}{3}.
\end{equation}
In other words, under these assumptions, the absence of domain wall-compatible
events would 
favour encounter rates producing less than (approximately) three expected events during the observation window 
and encounter time larger then a third of the observation window.
It should be stressed that a final upper-limit curve would require repeating
this recipe over a range of coincidence orders, injected signal amplitudes
and nominal searching-thresholds.
Each search channel $(n,\theta)$ defines its own background
expectation $b_{n,\theta}$ and detection efficiency $\epsilon_{n,\theta}(A)$,
and therefore its own upper limit $R_{95}^{(n,\theta)}(A)$.
The result of this procedure could then be
translated into a bound on the particle parameters of the model,
for example the electron-scalar effective scale $\Lambda_e$ - as also concluded by \textit{Roberts et. al.} in \cite{Roberts:2017hla}.

The mentioned approach or interpretation should be regarded as \emph{only indicative}. 
In the present case, the expected number of noise-induced events for the selected threshold 
$b_{n,\theta}$ is of the order of $10^{-2}$
which is higher than a (required) strong background-free regime where $b_{n,\theta} \leq 10^{-6}$.
The time series values, the stability analysis and the empirical false-alarm curves
-- already presented in Figures ~\ref{Fig8}, \ref{Fig9} and \ref{Fig10} --
indicate that the Galileo network is affected by correlations among different nodes and by non-stationary noise properties. 
As a result, the background can vary significantly over the available observation time. 
Therefore, a search strategy solely based on one 
or (possibly) more fixed-threshold channels $(n_i,\theta_i)$ must rely
on sufficiently high thresholds in order to suppress the contribution of particularly noisy periods. 
Nevertheless, this choice may also reduce the sensitivity
during less-noisy intervals, where a physical signal could otherwise be nestled.

In this respect, a more refined strategy would be to characterize the background locally in time
via some specific recipe, as a \emph{moving-average analysis}, for instance. 
For example, one may divide the data into
shorter temporal segments and assign to each segment its own mean and standard
deviation, or more generally a local estimate $(\mu_i,\sigma_i)$ of the noise level. 
This would allow one to define a local, or adaptive, threshold describing
the instantaneous state of the network, rather than imposing a single global
threshold over the full observation window.
In the GNSS context, such a local-background prescription could then be combined
with the mentioned topological pattern consistency checks to obtain more reliable and robust constraints.

This approach will be properly studied in future developments. 
The goal here was to show how the inclusion of FAR and detection-efficiency information improves the previous search recipe, in which
these ingredients were not explicitly included, while also assessing the
remaining limitations of a fixed-threshold strategy.
In this sense, this analysis
allowed to identify where the statistical framework must be refined and how a more robust search
pipeline should be developed.

\section{Conclusions and prospects}
\label{Conclusions and prospects}

The possibility of using an extended network of precision quantum sensors 
in order to catch up
possible domain wall-like crossings represents a relatively new interesting challenge.
In particular, using a GNSS atomic clock network would allow one to test the passage of objects
with very large size, typically generated by ultralight scalar or pseudoscalar fields in some early universe phase.
In this work, the focus is on the impact of Earth-sized ultralight scalar field
on the Galileo GNSS atomic clock network
taking into account the quality of the clock data.

Starting from a simple scalar-field model with a spontaneously
broken discrete symmetry at energy below the electroweak phase transition,
effective quadratic couplings to the electron and photon sectors were considered. 
In this framework, the passage of a domain wall can induce transient variations of electron mass and fine-structure constant,
leading to perturbations in the nominal frequencies of the GNSS atomic clocks
which, in turn, lead to a precise signature on the satellite atomic clock bias difference time series. 
In particular, a domain-wall crossing is expected to produce a characteristic pattern across the set of
bias derivative time-series 
made of a common \textit{positive spikes} (determining a coincidence event, involving all or a certain number of satellites), 
associated with the domain wall-reference clock interaction, 
together with \textit{negative spikes} distributed around the positive one and according to the propagation of the
wall across the GNSS network. 
The simplest version of this pattern is achieved if one assumes the scalar-electron coupling dominates 
the interaction physics.
The search for \textit{partial or complete coincidence events} provides the physical basis for testing
a domain wall crossing.

The search for such signatures necessarily requires a detailed characterization of the relevant time series. 
In this work, this characterization was carried out for a selected subnetwork of eleven Galileo
satellites over their common working periods in the available dataset. 
After removing residual trends and explicit discontinuities, the main features and stability
properties in time and frequency domain were investigated revealing a complex background with satellite-dependent noise,
residual correlations, local non-stationarities, and occasional large excursions.
The unavoidable sampling nature of the time series and the summarized strong noisy behavior
represent crucial aspects that one must take into account.

In particular, in order to quantify the probability that this background mimics a
domain-wall-like event, a false-alarm-rate analysis was performed using both a
theoretical Poisson-Binomial description and an empirical time-slide procedure.
This was complemented by an injection-based detection-efficiency study, in
which $\sim 10^3$ simulated domain-wall crossings were propagated through the
Galileo geospace and injected into the real bias-difference data. The combined
FAR and detection-efficiency information indicates that the $7$-fold coincidence,
with a searching-threshold of order $1.5\times 10^{-10}$ seconds, could represent a favorable
benchmark for a preliminary search of signals with larger amplitudes, providing a useful compromise between
background suppression and signal recovery.

In principle, this selected channel can be used, at least as an illustrative exercise,
to estimate the type of constraint that would follow from a hypothetical null result within
a simple Poisson counting framework at some confidence level, following the pioneering work presented in \cite{Roberts:2017hla}.
However, such fixed-threshold prescription cannot be regarded as fully robust
also considering standard FAR and detection efficiency outcomes.
Indeed, the observed correlations and
non-stationary properties of the Galileo background  (via the FAR analysis) require a more refined
characterization, in which the noise level should be estimated locally in time. 
From this point of view, a natural extension could be represented by a moving average analysis,
allowing one to define adaptive thresholds that follow 
the almost instantaneous noise-state of the network. This will be the focus of future extension(s) of the present work.

The analysis discussed in this work is based on the Galileo GNSS atomic-clock
data produced with the NAPEOS software within the GREAT project, as anticipated
in Sec.~\ref{GNSS clock data: Precise orbit determination, clock bias generation and sampling}.
Although a deeper characterization of these clock-bias data and of their impact
on a possible domain-wall measurement is clearly of interest, a crucial
step forward would consist of generating independent and more robust clock solutions.
This can be achieved, for example, with the Bernese GNSS Software, by
combining microwave GNSS observations with SLR data.
Such an approach can help to reduce the radial orbit errors, which significantly affect the estimated clock bias values.

In this respect, it would also be interesting to explore the impact of additional
effects, such as more general clock trajectories, nontrivial domain-wall
propagation velocities, or the use of data with sampling intervals shorter than $30$ seconds.
However, in the latter case, one should also change the search strategy since spike-like signatures are no longer a suitable benchmark.

As a final remark, a crucial point should be outlined.
In the presence of a hypothetical candidate or more favorable candidates, 
the microscopic interpretation would not be unique. 
A signal in the clock-bias or RFF data could certainly arise solely from the coupling to the electron sector,
under the assumptions adopted here.
However, it could also derive from additional contributions, like from (smaller) variations of the fine-structure constant
or from other Standard Model (Higgs, quarks) and beyond-Standard-Model (supersymmetric-like?) portals. 
Breaking this degeneracy would necessarily require complementary measurements with independent clock
species or additional platforms of precision-sensor networks.

\section{Acknowledgments}
This work is part of the G4S$\_$2.0 project, developed under the auspices of the Italian Space Agency (ASI) within
the frame of the Bando Premiale CI-COT-2018-085, under the Accordo Attuativo No. 2021-14-HH.0, with co-participation of
the Italian Institute for Astrophysics (INAF) and the Politecnico di Torino (POLITO).

\newcommand{\journal}[2]{\href{http://dx.doi.org/#1}{#2}}
\newcommand{\arxiv}[2]{\href{http://arxiv.org/abs/#1}{[arxiv:#1 #2]}}


\begin{thebibliography}{150}



\bibitem{Burgess:2006hbd}
C.~P.~Burgess and G.~D.~Moore,
``The standard model: A primer,''
Cambridge University Press, 2006,
\journal{doi.org/10.1017/CBO9780511819698}{Cambridge University Press, 2006}

\bibitem{Thomson:2013zua}
M.~Thomson,
``Modern particle physics,''
\journal{10.1017/CBO9781139525367}{Cambridge University Press, 2013}

\bibitem{Kane:2017aaa}
G.~Kane, 
``Modern Elementary Particle Physics: Explaining and Extending the Standard Model,''
\journal{doi.org/10.1017/9781316691434}{2nd ed. Cambridge: Cambridge University Press}

\bibitem{Donoghue:1992dd}
J.~F.~Donoghue, E.~Golowich and B.~R.~Holstein,
``Dynamics of the standard model,''
\journal{10.1017/CBO9780511524370}{Camb. Monogr. Part. Phys. Nucl. Phys. Cosmol. \textbf{2}, 1-540 (1992)}


\bibitem{Lee:1974jb}
T.~D.~Lee,
``CP Nonconservation and Spontaneous Symmetry Breaking,''
\journal{10.1016/0370-1573(74)90020-9}{Phys. Rept. \textbf{9}, 143-177 (1974)}


\bibitem{Zeldovich:1974uw}
Y.~B.~Zeldovich, I.~Y.~Kobzarev and L.~B.~Okun,
``Cosmological Consequences of the Spontaneous Breakdown of Discrete Symmetry,''
\journal{}{Zh. Eksp. Teor. Fiz. \textbf{67}, 3-11 (1974)}

\bibitem{Kibble:1976sj}
T.~W.~B.~Kibble,
``Topology of Cosmic Domains and Strings,''
\journal{10.1088/0305-4470/9/8/029}{J. Phys. A \textbf{9}, 1387-1398 (1976)}

\bibitem{Kibble:1980mv}
T.~W.~B.~Kibble,
``Some Implications of a Cosmological Phase Transition,''
\journal{10.1016/0370-1573(80)90091-5}{Phys. Rept. \textbf{67}, 183 (1980)}


\bibitem{Lazarides:1981fv}
G.~Lazarides, Q.~Shafi and T.~F.~Walsh,
``Cosmic Strings and Domains in Unified Theories,''
\journal{10.1016/0550-3213(82)90052-9}{Nucl. Phys. B \textbf{195}, 157-172 (1982)}


\bibitem{Vilenkin:2000jqa}
A.~Vilenkin and E.~P.~S.~Shellard,
``Cosmic Strings and Other Topological Defects,''
Cambridge University Press, 2000.
\journal{}{Cambridge University Press, 2000.}

\bibitem{Vachaspati:2006zz}
T.~Vachaspati,
``Kinks and Domain Walls : An Introduction to Classical and Quantum Solitons,''
\journal{10.1017/9781009290456}{Oxford University Press, 2007}



\bibitem{Vilenkin:1981zs}
A.~Vilenkin,
``Gravitational Field of Vacuum Domain Walls and Strings,''
\journal{10.1103/PhysRevD.23.852}{Phys. Rev. D \textbf{23}, 852-857 (1981)}


\bibitem{Vilenkin:1983ykn}
A.~Vilenkin,
``Gravitational Field of Vacuum Domain Walls,''
\journal{10.1016/0370-2693(83)90554-3}{Phys. Lett. B \textbf{133}, 177-179 (1983)}




\bibitem{Kibble:1982dd}
T.~W.~B.~Kibble, G.~Lazarides and Q.~Shafi,
``Walls Bounded by Strings,''
\journal{10.1103/PhysRevD.26.435}{Phys. Rev. D \textbf{26}, 435 (1982)}

\bibitem{Vilenkin:1982ks}
A.~Vilenkin and A.~E.~Everett,
``Cosmic Strings and Domain Walls in Models with Goldstone and PseudoGoldstone Bosons,''
\journal{10.1103/PhysRevLett.48.1867}{Phys. Rev. Lett. \textbf{48}, 1867-1870 (1982)}


\bibitem{Kuzmin:1981ip}
V.~A.~Kuzmin, M.~E.~Shaposhnikov and I.~I.~Tkachev,
``Matter - Antimatter Domains in the Universe: A Solution of the Vacuum Walls Problem,''
\journal{10.1016/0370-2693(81)91013-3}{Phys. Lett. B \textbf{105}, 167-170 (1981)}



\bibitem{Mohanty:1984pj}
A.~K.~Mohanty and F.~W.~Stecker,
``Matter - Antimatter Domains: A Possible Solution to the {CP} Domain Wall Problem in the Early Universe,''
\journal{10.1016/0370-2693(84)91481-3}{Phys. Lett. B \textbf{143}, 351-356 (1984)}

\bibitem{Tamvakis:1981aaa}
K.~Tamvakis and C.~E.~Vayonakis
``A possible solution to the problem of cosmic domain walls''
\journal{10.1016/0370-2693(82)90452-X}{Phys. Lett. B \textbf{109}, 15-18 (1982)}



\bibitem{Preskill:1991kd}
J.~Preskill, S.~P.~Trivedi, F.~Wilczek and M.~B.~Wise,
``Cosmology and broken discrete symmetry,''
\journal{10.1016/0550-3213(91)90241-O}{Nucl. Phys. B \textbf{363}, 207-220 (1991)}




\bibitem{Sikivie:1982qv}
P.~Sikivie,
``Of Axions, Domain Walls and the Early Universe,''
\journal{10.1103/PhysRevLett.48.1156}{Phys. Rev. Lett. \textbf{48}, 1156-1159 (1982)}



\bibitem{Rai:1992xw}
B.~Rai and G.~Senjanovic,
``Gravity and domain wall problem,''
\journal{10.1103/PhysRevD.49.2729}{Phys. Rev. D \textbf{49}, 2729-2733 (1994)}
\arxiv{9301240}{[hep-ph]}






\bibitem{Giddings:1987cg}
S.~B.~Giddings and A.~Strominger,
``Axion Induced Topology Change in Quantum Gravity and String Theory,''
\journal{10.1016/0550-3213(88)90446-4}{Nucl. Phys. B \textbf{306}, 890-907 (1988)}

\bibitem{Banks:1988yz}
T.~Banks and L.~J.~Dixon,
``Constraints on String Vacua with Space-Time Supersymmetry,''
\journal{10.1016/0550-3213(88)90523-8}{Nucl. Phys. B \textbf{307}, 93-108 (1988)}

\bibitem{Kallosh:1995hi}
R.~Kallosh, A.~D.~Linde, D.~A.~Linde and L.~Susskind,
``Gravity and global symmetries,''
\journal{10.1103/PhysRevD.52.912}{Phys. Rev. D \textbf{52}, 912-935 (1995)}
\arxiv{hep-th/9502069}{[hep-th]}


\bibitem{Banks:2010zn}
T.~Banks and N.~Seiberg,
``Symmetries and Strings in Field Theory and Gravity,''
\journal{10.1103/PhysRevD.83.084019}{Phys. Rev. D \textbf{83}, 084019 (2011)}
\arxiv{1011.5120}{[hep-th]}


\bibitem{Vafa:2005ui}
C.~Vafa,
``The String landscape and the swampland,''
\arxiv{hep-th/0509212}{[hep-th]}

\bibitem{Arkani-Hamed:2006emk}
N.~Arkani-Hamed, L.~Motl, A.~Nicolis and C.~Vafa,
``The String landscape, black holes and gravity as the weakest force,''
\journal{10.1088/1126-6708/2007/06/060}{JHEP \textbf{06}, 060 (2007)}
\arxiv{0601001}{[hep-th]}

\bibitem{Ooguri:2006in}
H.~Ooguri and C.~Vafa,
``On the Geometry of the String Landscape and the Swampland,''
\journal{10.1016/j.nuclphysb.2006.10.033}{Nucl. Phys. B \textbf{766}, 21-33 (2007)}
\arxiv{hep-th/0605264}{[hep-th]}

\bibitem{Fichet:2019ugl}
S.~Fichet and P.~Saraswat,
``Approximate Symmetries and Gravity,''
\journal{10.1007/JHEP01(2020)088}{JHEP \textbf{01}, 088 (2020)}
\arxiv{1909.02002}{[hep-th]}


\bibitem{Harlow:2018tng}
D.~Harlow and H.~Ooguri,
``Symmetries in quantum field theory and quantum gravity,''
\journal{10.1007/s00220-021-04040-y}{Commun. Math. Phys. \textbf{383}, no.3, 1669-1804 (2021)}
\arxiv{1810.05338}{[hep-th]}


\bibitem{Daus:2020vtf}
T.~Daus, A.~Hebecker, S.~Leonhardt and J.~March-Russell,
``Towards a Swampland Global Symmetry Conjecture using weak gravity,''
\journal{10.1016/j.nuclphysb.2020.115167}{Nucl. Phys. B \textbf{960}, 115167 (2020)}
\arxiv{2002.02456}{[hep-th]}




\bibitem{Press:1989yh}
W.~H.~Press, B.~S.~Ryden and D.~N.~Spergel,
``Dynamical Evolution of Domain Walls in an Expanding Universe,''
\journal{10.1086/168151}{Astrophys. J. \textbf{347}, 590-604 (1989)}


\bibitem{Gelmini:1988sf}
G.~B.~Gelmini, M.~Gleiser and E.~W.~Kolb,
``Cosmology of Biased Discrete Symmetry Breaking,''
\journal{10.1103/PhysRevD.39.1558}{Phys. Rev. D \textbf{39}, 1558 (1989)}

\bibitem{Coulson:1995nv}
D.~Coulson, Z.~Lalak and B.~A.~Ovrut,
``Biased domain walls,''
\journal{10.1103/PhysRevD.53.4237}{Phys. Rev. D \textbf{53}, 4237-4246 (1996)}

\bibitem{Larsson:1996sp}
S.~E.~Larsson, S.~Sarkar and P.~L.~White,
``Evading the cosmological domain wall problem,''
\journal{10.1103/PhysRevD.55.5129}{Phys. Rev. D \textbf{55}, 5129-5135 (1997)}
\arxiv{hep-ph/9608319}{[hep-ph]}





\bibitem{Hindmarsh:1996xv}
M.~Hindmarsh,
``Analytic scaling solutions for cosmic domain walls,''
\journal{10.1103/PhysRevLett.77.4495}{Phys. Rev. Lett. \textbf{77}, 4495-4498 (1996)}
\arxiv{hep-ph/9605332}{[hep-ph]}


\bibitem{Hindmarsh:2002bq}
M.~Hindmarsh,
``Level set method for the evolution of defect and brane networks,''
\journal{10.1103/PhysRevD.68.043510}{Phys. Rev. D \textbf{68}, 043510 (2003)}
\arxiv{hep-ph/0207267}{[hep-ph]}

\bibitem{Avelino:2005kn}
P.~P.~Avelino, C.~J.~A.~P.~Martins and J.~C.~R.~E.~Oliveira,
``One-scale model for domain wall network evolution,''
\journal{10.1103/PhysRevD.72.083506}{Phys. Rev. D \textbf{72}, 083506 (2005)}
\arxiv{hep-ph/0507272}{[hep-ph]}



\bibitem{Martins:2016lzc}
C.~J.~A.~P.~Martins, I.~Y.~Rybak, A.~Avgoustidis and E.~P.~S.~Shellard,
``Stretching and Kibble scaling regimes for Hubble-damped defect networks,''
\journal{10.1103/PhysRevD.94.116017}{Phys. Rev. D \textbf{94}, no.11, 116017 (2016) [erratum: Phys. Rev. D \textbf{95}, no.3, 039902 (2017)}
\arxiv{1612.08863}{[hep-ph]}

\bibitem{Avelino:2019wqd}
P.~P.~Avelino,
``Parameter-free velocity-dependent one-scale model for domain walls,''
\journal{10.1103/PhysRevD.101.023514}{Phys. Rev. D \textbf{101}, no.2, 023514 (2020)}
\arxiv{1910.07011}{[astro-ph.CO]}

\bibitem{Avelino:2020ubr}
P.~P.~Avelino,
``Comparing parametric and non-parametric velocity-dependent one-scale models for domain wall evolution,''
\journal{10.1088/1475-7516/2020/04/012}{JCAP \textbf{04}, 012 (2020)}
\arxiv{2001.06318}{[astro-ph.CO]}

\bibitem{Avelino:2022zem}
P.~Avelino, D.~Gr{\"u}ber and L.~Sousa,
``Analytical Scaling Solutions for the Evolution of Cosmic Domain Walls in a Parameter-Free Velocity-Dependent One-Scale Model,''
\journal{10.3390/sym14091799}{Symmetry \textbf{14}, no.9, 1799 (2022)}
\arxiv{2203.16173}{[astro-ph.CO]}



\bibitem{Garagounis:2002kt}
T.~Garagounis and M.~Hindmarsh,
``Scaling in numerical simulations of domain walls,''
\journal{10.1103/PhysRevD.68.103506}{Phys. Rev. D \textbf{68}, 103506 (2003)}
\arxiv{hep-ph/0212359}{[hep-ph]}

\bibitem{Oliveira:2004he}
J.~C.~R.~E.~Oliveira, C.~J.~A.~P.~Martins and P.~P.~Avelino,
``The Cosmological evolution of domain wall networks,''
\journal{10.1103/PhysRevD.71.083509}{Phys. Rev. D \textbf{71}, 083509 (2005)}
\arxiv{hep-ph/0410356}{[hep-ph]}

\bibitem{Avelino:2005pe}
P.~P.~Avelino, J.~C.~R.~E.~Oliveira and C.~J.~A.~P.~Martins,
``Understanding domain wall network evolution,''
\journal{10.1016/j.physletb.2005.02.003}{Phys. Lett. B \textbf{610}, 1-8 (2005)}
\arxiv{hep-th/0503226}{[hep-th]}

\bibitem{Leite:2011sc}
A.~M.~M.~Leite and C.~J.~A.~P.~Martins,
``Scaling Properties of Domain Wall Networks,''
\journal{10.1103/PhysRevD.84.103523}{Phys. Rev. D \textbf{84}, 103523 (2011)}
\arxiv{1110.3486}{[hep-ph]}

\bibitem{Leite:2012vn}
A.~M.~M.~Leite, C.~J.~A.~P.~Martins and E.~P.~S.~Shellard,
``Accurate Calibration of the Velocity-dependent One-scale Model for Domain Walls,''
\journal{10.1016/j.physletb.2012.11.070}{Phys. Lett. B \textbf{718}, 740-744 (2013)}
\arxiv{1206.6043}{[hep-ph]}

\bibitem{Correia:2014kqa}
J.~R.~C.~C.~C.~Correia, I.~S.~C.~R.~Leite and C.~J.~A.~P.~Martins,
``Effects of Biases in Domain Wall Network Evolution,''
\journal{10.1103/PhysRevD.90.023521}{Phys. Rev. D \textbf{90}, no.2, 023521 (2014)}
\arxiv{1407.3905}{[hep-ph]}


\bibitem{Martins:2016ois}
C.~J.~A.~P.~Martins, I.~Y.~Rybak, A.~Avgoustidis and E.~P.~S.~Shellard,
``Extending the velocity-dependent one-scale model for domain walls,''
\journal{10.1103/PhysRevD.93.043534}{Phys. Rev. D \textbf{93}, no.4, 043534 (2016)}
\arxiv{1602.01322}{[hep-ph]}

\bibitem{Correia:2018tty}
J.~R.~C.~C.~C.~Correia, I.~S.~C.~R.~Leite and C.~J.~A.~P.~Martins,
``Effects of biases in domain wall network evolution. II. Quantitative analysis,''
\journal{10.1103/PhysRevD.97.083521}{Phys. Rev. D \textbf{97}, no.8, 083521 (2018)}
\arxiv{1804.10761}{[astro-ph.CO]}











\bibitem{Bucher:1998mh}
M.~Bucher and D.~N.~Spergel,
``Is the dark matter a solid?,''
\journal{10.1103/PhysRevD.60.043505}{Phys. Rev. D \textbf{60}, 043505 (1999)}
\arxiv{astro-ph/9812022}{[astro-ph]}


\bibitem{Battye:1999eq}
R.~A.~Battye, M.~Bucher and D.~Spergel,
``Domain wall dominated universes,''
\arxiv{astro-ph/9908047}{[astro-ph]}


\bibitem{Friedland:2002qs}
A.~Friedland, H.~Murayama and M.~Perelstein,
``Domain walls as dark energy,''
\journal{10.1103/PhysRevD.67.043519}{Phys. Rev. D \textbf{67}, 043519 (2003)}
\arxiv{astro-ph/0205520}{[astro-ph]}



\bibitem{PinaAvelino:2006ia}
P.~Pina Avelino, C.~J.~A.~P.~Martins, J.~Menezes, R.~Menezes and J.~C.~R.~E.~Oliveira,
``Frustrated expectations: defect networks and dark energy,''
\journal{10.1103/PhysRevD.73.123519}{Phys. Rev. D \textbf{73}, 123519 (2006)}
\arxiv{astro-ph/0602540}{[astro-ph]}

\bibitem{Avelino:2006xy}
P.~P.~Avelino, C.~J.~A.~P.~Martins, J.~Menezes, R.~Menezes and J.~C.~R.~E.~Oliveira,
``Defect junctions and domain wall dynamics,''
\journal{10.1103/PhysRevD.73.123520}{Phys. Rev. D \textbf{73}, 123520 (2006)}
\arxiv{hep-ph/0604250}{[hep-ph]}

\bibitem{Avelino:2008ve}
P.~P.~Avelino, C.~J.~A.~P.~Martins, J.~Menezes, R.~Menezes and J.~C.~R.~E.~Oliveira,
``Dynamics of domain wall networks with junctions,''
\journal{10.1103/PhysRevD.78.103508}{Phys. Rev. D \textbf{78}, 103508 (2008)}
\arxiv{0807.4442}{[hep-ph]}

\bibitem{Avelino:2008qy}
P.~P.~Avelino, C.~J.~A.~P.~Martins and L.~Sousa,
``Dynamics of Biased Domain Walls and the Devaluation Mechanism,''
\journal{10.1103/PhysRevD.78.043521}{Phys. Rev. D \textbf{78}, 043521 (2008)}
\arxiv{0805.4013}{[astro-ph]}


\bibitem{Sousa:2015cqa}
L.~Sousa and P.~P.~Avelino,
``Cosmic Microwave Background anisotropies generated by domain wall networks,''
\journal{10.1103/PhysRevD.92.083520}{Phys. Rev. D \textbf{92}, no.8, 083520 (2015)}
\arxiv{1507.01064}{[astro-ph.CO]}


\bibitem{Lazanu:2015fua}
A.~Lazanu, C.~J.~A.~P.~Martins and E.~P.~S.~Shellard,
``Contribution of domain wall networks to the CMB power spectrum,''
\journal{10.1016/j.physletb.2015.06.034}{Phys. Lett. B \textbf{747}, 426-432 (2015)}
\arxiv{1505.03673}{[astro-ph.CO]}




\bibitem{Takahashi:2020tqv}
F.~Takahashi and W.~Yin,
``Kilobyte Cosmic Birefringence from ALP Domain Walls,''
\journal{10.1088/1475-7516/2021/04/007}{JCAP \textbf{04}, 007 (2021)}
\arxiv{2012.11576}{[hep-ph]}

\bibitem{Fujita:2020ecn}
T.~Fujita, K.~Murai, H.~Nakatsuka and S.~Tsujikawa,
``Detection of isotropic cosmic birefringence and its implications for axionlike particles including dark energy,''
\journal{10.1103/PhysRevD.103.043509}{Phys. Rev. D \textbf{103}, no.4, 043509 (2021)}
\arxiv{2011.11894}{[astro-ph.CO]}

\bibitem{Gonzalez:2022mcx}
D.~Gonzalez, N.~Kitajima, F.~Takahashi and W.~Yin,
``Stability of domain wall network with initial inflationary fluctuations and its implications for cosmic birefringence,''
\journal{10.1016/j.physletb.2023.137990}{Phys. Lett. B \textbf{843}, 137990 (2023)}
\arxiv{2211.06849}{[hep-ph]}

\bibitem{Ferreira:2023jbu}
R.~Z.~Ferreira, S.~Gasparotto, T.~Hiramatsu, I.~Obata and O.~Pujolas,
``Axionic defects in the CMB: birefringence and gravitational waves,''
\journal{10.1088/1475-7516/2024/05/066}{JCAP \textbf{05}, 066 (2024)}
\arxiv{2312.14104}{[hep-ph]}


\bibitem{Lee:2025yvn}
J.~Lee, K.~Murai, F.~Takahashi and W.~Yin,
``Isotropic cosmic birefringence from string axion domain walls without cosmic strings and DESI results,''
\journal{10.1103/pjf4-sn4v}{Phys. Rev. D \textbf{112}, no.4, 043538 (2025)}
\arxiv{2503.18417}{[hep-ph]}



\bibitem{Brandenberger:1994mq}
R.~H.~Brandenberger, A.~C.~Davis, T.~Prokopec and M.~Trodden,
``Local and nonlocal defect mediated electroweak baryogenesis,''
\journal{10.1103/PhysRevD.53.4257}{Phys. Rev. D \textbf{53}, 4257-4266 (1996)}
\arxiv{9409281}{[hep-ph]}

\bibitem{Abel:1995uc}
S.~A.~Abel and P.~L.~White,
``Baryogenesis from domain walls in the next-to-minimal supersymmetric standard model,''
\journal{10.1103/PhysRevD.52.4371}{Phys. Rev. D \textbf{52}, 4371-4379 (1995)}
\arxiv{9505241}{[hep-ph]}

\bibitem{Prokopec:1995gp}
T.~Prokopec, R.~H.~Brandenberger, A.~C.~Davis and M.~Trodden,
``Dynamical breaking of CPT symmetry in defect networks and baryogenesis,''
\journal{10.1016/0370-2693(96)00602-8}{Phys. Lett. B \textbf{384}, 175-179 (1996)}
\arxiv{9511349}{[hep-ph]}



\bibitem{Brandenberger:1998kn}
R.~H.~Brandenberger, I.~E.~Halperin and A.~Zhitnitsky,
``Baryogenesis with QCD domain walls,''
\arxiv{9808471}{[hep-ph]}

\bibitem{Daido:2015gqa}
R.~Daido, N.~Kitajima and F.~Takahashi,
``Axion domain wall baryogenesis,''
\journal{10.1088/1475-7516/2015/07/046}{JCAP \textbf{07}, 046 (2015)}
\arxiv{1504.07917}{[hep-ph]}

\bibitem{Daido:2015bva}
R.~Daido, N.~Kitajima and F.~Takahashi,
``Domain Wall Formation from Level Crossing in the Axiverse,''
\journal{10.1103/PhysRevD.92.063512}{Phys. Rev. D \textbf{92}, no.6, 063512 (2015)}
\arxiv{1505.07670}{[hep-ph]}


\bibitem{Schroder:2024gsi}
T.~Schr{\"o}der and R.~Brandenberger,
``Embedded domain walls and electroweak baryogenesis,''
\journal{10.1103/PhysRevD.110.043516}{Phys. Rev. D \textbf{110}, no.4, 043516 (2024)}
\arxiv{2404.13035}{[hep-ph]}



\bibitem{Mariotti:2024eoh}
A.~Mariotti, X.~Nagels, A.~Rase and M.~Vanvlasselaer,
``DW-genesis: baryon number from domain wall network collapse,''
\journal{10.1007/JHEP03(2025)199}{JHEP \textbf{03}, 199 (2025)}
\arxiv{2411.13494}{[hep-ph]}


\bibitem{Azzola:2024pzq}
J.~Azzola, O.~Matsedonskyi and A.~Weiler,
``Minimal electroweak baryogenesis via domain walls,''
\journal{10.1007/JHEP04(2025)103}{JHEP \textbf{04}, 103 (2025)}
\arxiv{2412.10495}{[hep-ph]}

\bibitem{Azzola:2026cwa}
J.~Azzola, O.~Matsedonskyi and A.~Weiler,
``Baryon Asymmetry from Electroweak-Symmetric Domain Walls,''
\arxiv{2604.16603}{[hep-ph]}



\bibitem{Stadnik:2014cea}
Y.~V.~Stadnik and V.~V.~Flambaum,
``Searches for topological defect dark matter via nongravitational signatures,''
\journal{10.1103/PhysRevLett.113.151301}{Phys. Rev. Lett. \textbf{113}, no.15, 151301 (2014)}
\arxiv{1405.5337}{[hep-ph]}


\bibitem{Naik:2022lcn}
A.~P.~Naik and C.~Burrage,
``Dark sector domain walls could explain the observed planes of satellites,''
\journal{10.1088/1475-7516/2022/08/020}{JCAP \textbf{08}, no.08, 020 (2022)}
\arxiv{2205.00712}{[astro-ph.GA]}





\bibitem{Chang:1998tb}
S.~Chang, C.~Hagmann and P.~Sikivie,
``Studies of the motion and decay of axion walls bounded by strings,''
\journal{10.1103/PhysRevD.59.023505}{Phys. Rev. D \textbf{59}, 023505 (1999)}
\arxiv{hep-ph/9807374}{[hep-ph]}

\bibitem{Gleiser:1998na}
M.~Gleiser and R.~Roberts,
``Gravitational waves from collapsing vacuum domains,''
\journal{10.1103/PhysRevLett.81.5497}{Phys. Rev. Lett. \textbf{81}, 5497-5500 (1998)}
\arxiv{astro-ph/9807260}{[astro-ph]}

\bibitem{Hiramatsu:2010yz}
T.~Hiramatsu, M.~Kawasaki and K.~Saikawa,
``Gravitational Waves from Collapsing Domain Walls,''
\journal{10.1088/1475-7516/2010/05/032}{JCAP \textbf{05}, 032 (2010)}
\arxiv{1002.1555}{[astro-ph.CO]}

\bibitem{Kawasaki:2011vv}
M.~Kawasaki and K.~Saikawa,
``Study of gravitational radiation from cosmic domain walls,''
\journal{10.1088/1475-7516/2011/09/008}{JCAP \textbf{09}, 008 (2011)}
\arxiv{1102.5628}{[astro-ph.CO]}



\bibitem{Saikawa:2017hiv}
K.~Saikawa,
``A review of gravitational waves from cosmic domain walls,''
\journal{10.3390/universe3020040}{Universe \textbf{3}, no.2, 40 (2017)}
\arxiv{1703.02576}{[hep-ph]}


\bibitem{Hiramatsu:2013qaa}
T.~Hiramatsu, M.~Kawasaki and K.~Saikawa,
``On the estimation of gravitational wave spectrum from cosmic domain walls,''
\journal{10.1088/1475-7516/2014/02/031}{JCAP \textbf{02}, 031 (2014)}
\arxiv{1309.5001}{[astro-ph.CO]}

\bibitem{Roshan:2024qnv}
R.~Roshan and G.~White,
``Using gravitational waves to see the first second of the Universe,''
\journal{10.1103/RevModPhys.97.015001}{Rev. Mod. Phys. \textbf{97}, no.1, 015001 (2025)}
\arxiv{2401.04388}{[hep-ph]}

\bibitem{Blasi:2025tmn}
S.~Blasi, A.~Mariotti, A.~Rase and M.~Vanvlasselaer,
``Domain walls in the scaling regime: Equal Time Correlator and Gravitational Waves,''
\arxiv{2511.16649}{[hep-ph]}



\bibitem{King:2023ztb}
S.~F.~King, R.~Roshan, X.~Wang, G.~White and M.~Yamazaki,
``Quantum gravity effects on fermionic dark matter and gravitational waves,''
\journal{10.1088/1475-7516/2024/05/071}{JCAP \textbf{05}, 071 (2024)}
\arxiv{2311.12487}{[hep-ph]}

\bibitem{King:2023ayw}
S.~F.~King, R.~Roshan, X.~Wang, G.~White and M.~Yamazaki,
``Quantum gravity effects on dark matter and gravitational waves,''
\journal{10.1103/PhysRevD.109.024057}{Phys. Rev. D \textbf{109}, no.2, 024057 (2024)}
\arxiv{2308.03724}{[hep-ph]}

\bibitem{Borah:2024kfn}
D.~Borah, N.~Das and R.~Roshan,
``Observable gravitational waves and {\ensuremath{\Delta}}Neff with global lepton number symmetry and dark matter,''
\journal{10.1103/PhysRevD.110.075042}{Phys. Rev. D \textbf{110}, no.7, 075042 (2024)}
\arxiv{2406.04404}{[hep-ph]}




\bibitem{Ferrer:2018uiu}
F.~Ferrer, E.~Masso, G.~Panico, O.~Pujolas and F.~Rompineve,
``Primordial Black Holes from the QCD axion,''
\journal{10.1103/PhysRevLett.122.101301}{Phys. Rev. Lett. \textbf{122}, no.10, 101301 (2019)}
\arxiv{1807.01707}{[hep-ph]}

\bibitem{Gelmini:2022nim}
G.~B.~Gelmini, A.~Simpson and E.~Vitagliano,
``Catastrogenesis: DM, GWs, and PBHs from ALP string-wall networks,''
\journal{10.1088/1475-7516/2023/02/031}{JCAP \textbf{02}, 031 (2023)}
\arxiv{2207.07126}{[hep-ph]}

\bibitem{Gelmini:2023ngs}
G.~B.~Gelmini, J.~Hyman, A.~Simpson and E.~Vitagliano,
``Primordial black hole dark matter from catastrogenesis with unstable pseudo-Goldstone bosons,''
\journal{10.1088/1475-7516/2023/06/055}{JCAP \textbf{06}, 055 (2023)}
\arxiv{2303.14107}{[hep-ph]}


\bibitem{Dunsky:2024zdo}
D.~I.~Dunsky and M.~Kongsore,
``Primordial black holes from axion domain wall collapse,''
\journal{10.1007/JHEP06(2024)198}{JHEP \textbf{06}, 198 (2024)}
\arxiv{2402.03426}{[hep-ph]}



\bibitem{Gouttenoire:2023gbn}
Y.~Gouttenoire and E.~Vitagliano,
``Primordial black holes and wormholes from domain wall networks,''
\journal{10.1103/PhysRevD.109.123507}{Phys. Rev. D \textbf{109}, no.12, 123507 (2024)}
\arxiv{2311.07670}{[hep-ph]}



\bibitem{Gouttenoire:2023ftk}
Y.~Gouttenoire and E.~Vitagliano,
``Domain wall interpretation of the PTA signal confronting black hole overproduction,''
\journal{10.1103/PhysRevD.110.L061306}{Phys. Rev. D \textbf{110}, no.6, L061306 (2024)}
\arxiv{2306.17841}{[gr-qc]}

\bibitem{Ferreira:2024eru}
R.~Z.~Ferreira, A.~Notari, O.~Pujol{\`a}s and F.~Rompineve,
``Collapsing domain wall networks: impact on pulsar timing arrays and primordial black holes,''
\journal{10.1088/1475-7516/2024/06/020}{JCAP \textbf{06}, 020 (2024)}
\arxiv{2401.14331}{[astro-ph.CO]}


\bibitem{Gouttenoire:2025ofv}
Y.~Gouttenoire, S.~F.~King, R.~Roshan, X.~Wang, G.~White and M.~Yamazaki,
``Cosmological Consequences of Domain Walls Biased by Quantum Gravity,''
\arxiv{2501.16414}{[hep-ph]}






\bibitem{Pospelov:2012mt}
M.~Pospelov, S.~Pustelny, M.~P.~Ledbetter, D.~F.~Jackson Kimball, W.~Gawlik and D.~Budker,
``Detecting Domain Walls of Axionlike Models Using Terrestrial Experiments,''
\journal{10.1103/PhysRevLett.110.021803}{Phys. Rev. Lett. \textbf{110}, no.2, 021803 (2013)}
\arxiv{1205.6260}{[hep-ph]}

\bibitem{Pustelny:2013vza}
S.~Pustelny, D.~F.~Jackson Kimball, C.~Pankow, M.~P.~Ledbetter, P.~Wlodarczyk, P.~Wcislo, M.~Pospelov, J.~R.~Smith, J.~Read, W.~Gawlik and D.~Budker,
``The Global Network of Optical Magnetometers for Exotic physics (GNOME): A novel scheme to search for physics beyond the Standard Model,''
\journal{10.1002/andp.201300061}{Annalen Phys. \textbf{525}, no.8-9, 659-670 (2013)}
\arxiv{1303.5524}{[physics.atom-ph]}


\bibitem{Afach:2018eze}
S.~Afach, D.~Budker, G.~Decamp, V.~Dumont, Z.~D.~Gruji{\'c}, H.~Guo, D.~F.~Jackson Kimball, T.~W.~Kornack, V.~Lebedev and W.~Li, \textit{et al.}
``Characterization of the Global Network of Optical Magnetometers to search for Exotic Physics (GNOME),''
\journal{10.1016/j.dark.2018.10.002}{Phys. Dark Univ. \textbf{22}, 162-180 (2018)}
\arxiv{1807.09391}{[physics.ins-det]}

\bibitem{Masia-Roig:2019hsy}
H.~Masia-Roig, J.~A.~Smiga, D.~Budker, V.~Dumont, Z.~Grujic, D.~Kim, D.~F.~Jackson Kimball, V.~Lebedev, M.~Monroy and S.~Pustelny, \textit{et al.}
``Analysis method for detecting topological defect dark matter with a global magnetometer network,''
\journal{10.1016/j.dark.2020.100494}{Phys. Dark Univ. \textbf{28}, 100494 (2020)}
\arxiv{1912.08727}{[astro-ph.IM]}

\bibitem{GNOME:2023rpz}
S.~Afach \textit{et al.} [GNOME],
``What Can a GNOME Do?~ Search Targets for the Global Network of Optical Magnetometers for Exotic Physics Searches,''
\journal{10.1002/andp.202300083}{Annalen Phys. \textbf{536}, no.1, 2300083 (2024)}
\arxiv{2305.01785}{[hep-ph]}


\bibitem{McNally:2019lcg}
R.~L.~McNally and T.~Zelevinsky,
``Constraining domain wall dark matter with a network of superconducting gravimeters and LIGO,''
\journal{10.1140/epjd/e2020-100632-0}{Eur. Phys. J. D \textbf{74}, no.4, 61 (2020)}
\arxiv{1912.06703}{[astro-ph.IM]}




\bibitem{Jaeckel:2020mqa}
J.~Jaeckel, S.~Schenk and M.~Spannowsky,
``Probing dark matter clumps, strings and domain walls with gravitational wave detectors,''
\journal{10.1140/epjc/s10052-021-09604-9}{Eur. Phys. J. C \textbf{81}, no.9, 828 (2021)}
\arxiv{2004.13724}{[astro-ph.CO]}

\bibitem{Alvarez:2025frb}
E.~Alvarez, S.~Perkins, F.~Ravanedo and N.~Yunes,
``Dark matter clumps as sources of gravitational-wave glitches in LIGO-Virgo-KAGRA data,''
\journal{10.1103/jwbc-s8j1}{Phys. Rev. D \textbf{113}, no.4, 043058 (2026)}
\arxiv{2507.12227}{[gr-qc]}

\bibitem{Heisenberg:2023urf}
L.~Heisenberg, D.~Maibach and D.~Veske,
``Searching for topological dark matter in LIGO data,''
\journal{10.1103/PhysRevD.110.055037}{Phys. Rev. D \textbf{110}, no.5, 5 (2024)}
\arxiv{2309.05093}{[gr-qc]}

\bibitem{Thoss:2025yht}
V.~Thoss and A.~Loeb,
``Detecting dark objects in the Solar System with gravitational wave observatories,''
\journal{10.1103/g3wh-hg4x}{Phys. Rev. D \textbf{112}, no.8, 083050 (2025)}
\arxiv{2507.19577}{[gr-qc]}





\bibitem{Wcislo:2016}
P.~Wcis{\l}o, P.~Morzy{\'n}ski, M.~Bober, A.~Cygan, D.~Lisak, R.~Ciury{\l}o and M.~Zawada,
``Experimental constraint on dark matter detection with optical atomic clocks,''
\journal{10.1038/s41550-016-0009}{Nature Astron. \textbf{1}, 0009 (2016)}
\arxiv{1605.05763}{}


\bibitem{Wcislo:2018ojh}
P.~Wcis{\l}o, P.~Ablewski, K.~Beloy, S.~Bilicki, M.~Bober, R.~Brown, R.~Fasano, R.~Ciury{\l}o, H.~Hachisu and T.~Ido, \textit{et al.}
``New bounds on dark matter coupling from a global network of optical atomic clocks,''
\journal{10.1126/sciadv.aau4869}{Sci. Adv. \textbf{4}, no.12, eaau4869 (2018)}
\arxiv{1806.04762}{[physics.atom-ph]}


\bibitem{Roberts:2019sfo}
B.~M.~Roberts, P.~Delva, A.~Al-Masoudi, A.~Amy-Klein, C.~Baerentsen, C.~F.~A.~Baynham, E.~Benkler, S.~Bilicki, S.~Bize and W.~Bowden, \textit{et al.}
``Search for transient variations of the fine structure constant and dark matter using fiber-linked optical atomic clocks,''
\journal{10.1088/1367-2630/abaace}{New J. Phys. \textbf{22}, no.9, 093010 (2020}
\arxiv{1907.02661}{[astro-ph.CO]}





\bibitem{Derevianko:2013oaa}
A.~Derevianko and M.~Pospelov,
``Hunting for topological dark matter with atomic clocks,''
\journal{10.1038/nphys3137}{Nature Phys. \textbf{10}, 933 (2014)}
\arxiv{1311.1244}{[physics.atom-ph]}

\bibitem{Roberts:2017hla}
B.~M.~Roberts, G.~Blewitt, C.~Dailey, M.~Murphy, M.~Pospelov, A.~Rollings, J.~Sherman, W.~Williams and A.~Derevianko,
``Search for domain wall dark matter with atomic clocks on board global positioning system satellites,''
\journal{10.1038/s41467-017-01440-4}{Nature Commun. \textbf{8}, no.1, 1195 (2017)}
\arxiv{1704.06844}{[hep-ph]}

\bibitem{Roberts:2018xqn}
B.~M.~Roberts, G.~Blewitt, C.~Dailey and A.~Derevianko,
``Search for transient ultralight dark matter signatures with networks of precision measurement devices using a Bayesian statistics method,''
\journal{10.1103/PhysRevD.97.083009}{Phys. Rev. D \textbf{97}, no.8, 083009 (2018)}
\arxiv{1803.10264}{[astro-ph.IM]}

\bibitem{Roberts:2018agv}
B.~M.~Roberts and A.~Derevianko,
``Precision measurement noise asymmetry and its annual modulation as a dark matter signature,''
\journal{10.3390/universe7030050}{Universe \textbf{7}, no.3, 50 (2021)}
\arxiv{1803.00617}{[physics.atom-ph]}

\bibitem{Panelli:2019qtl}
G.~Panelli, B.~M.~Roberts and A.~Derevianko,
``Applying the matched-filter technique to the search for dark matter transients with networks of quantum sensors,''
\journal{10.1140/epjqt/s40507-020-00081-9}{EPJ Quantum Technol. \textbf{7}, no.1, 5 (2020)}
\arxiv{1908.03320}{[astro-ph.IM]}


\bibitem{Bertrand:2024ypu}
B.~Bertrand, P.~Defraigne, A.~Hees, A.~Sheremet, C.~Courde, J.~Chab{\'e}, J.~Ventura-Traveset, F.~Dilssner, E.~Schoenemann and L.~Mendes, \textit{et al.}
``Searching for large dark matter clumps using the Galileo Constellation clock variations,''
\journal{10.1016/j.asr.2024.03.080}{Adv. Space Res. \textbf{74}, 2551-2563 (2024)}
\arxiv{2403.17890}{[astro-ph.CO]}







\bibitem{Lewis:1991FrequencyStandards}
L.~L.~Lewis,
``An introduction to frequency standards,''
\journal{10.1109/5.84969}{Proc. IEEE \textbf{79}, no.7, 927--935 (1991)}




\bibitem{Montenbruck:2000satellite}
O.~Montenbruck and E.~Gill,
``Satellite Orbits: Models, Methods and Applications,''
\journal{10.1007/978-3-642-58351-3}{Springer-Verlag Berlin Heidelberg (2000)}



\bibitem{Lucchesi:2024zck}
D.~Lucchesi, M.~Visco, C.~Lefevre, M.~Lucente, F.~Santoli, F.~Sapio, M.~Cinelli, A.~Di Marco, E.~Fiorenza and P.~Loffredo, \textit{et al.}
``Fundamental physics measurements with Galileo FOC satellites and the Galileo for science project. I. A 3D-CAD and a box wing for modeling the effects of nonconservative forces,''
\journal{10.1103/PhysRevD.109.062004}{Phys. Rev. D \textbf{109}, no.6, 062004 (2024)}

\bibitem{Sapio:2024pwu}
F.~Sapio, D.~Lucchesi, M.~Visco, R.~Peron, M.~Lucente, C.~Lefevre, M.~Cinelli, A.~Di Marco, E.~Fiorenza and P.~Loffredo, \textit{et al.}
``Fundamental physics measurements with Galileo FOC satellites and the Galileo for science project. II. A box wing for modeling direct solar radiation pressure and preliminaries orbit determinations,''
\journal{10.1103/PhysRevD.109.062005}{Phys. Rev. D \textbf{109}, no.6, 062005 (2024)}


\bibitem{Petit:2010iers}
G.~Petit and B.~Luzum, eds.,
``IERS Conventions (2010),''
\journal{https://iers.org}{IERS Tech. Note \textbf{36}, 1-179 (2010)}


\bibitem{Lucchesi:2026lrs}
D.~Lucchesi,
``The scientific challenges of laser-ranged satellites,''
\journal{10.1007/s12210-025-01395-5}{Rend. Lincei Sci. Fis. Nat. (2026)}


\bibitem{dach2015bernese}
Dach, R., Lutz, S., Walser, P., \& Fridez, P. (Eds.). (2015).
\newblock {\em Bernese GNSS Software Version 5.2. User manual}.
\newblock Astronomical Institute, University of Bern.

\bibitem{mullins1972geodyn}
Mullins, N. E., Goad, C. C., Dao, N. C., Martin, T. V., Boulware, N. L., \& Chin, M. M. (1972). 
\textit{GEODYN programmers guide, volume 2, part 1} (No. NASA-CR-139150). 
NASA Contractor Report.

\bibitem{springer2011napeos}
Springer, T., Dil\ss{}ner, F., Escobar, D., Svehla, D., Flohrer, C., \& Zandbergen, R. (2011).
\newblock NAPEOS: The ESA/ESOC tool for space geodesy.
\newblock In {\em EGU General Assembly Conference Abstracts}, April 3-8, Vienna, Austria.

\bibitem{dow2009napeos}
Dow, J., \& Springer, T. (2009).
\newblock {\em NAPEOS Mathematical Models and Algorithms}.
\newblock DOPS-SYS-TN-0100-OPS-GN Issue 1.0, ESOC, Germany.



\bibitem{Delva:2018ilu}
P.~Delva, N.~Puchades, E.~Sch{\"o}nemann, F.~Dilssner, C.~Courde, S.~Bertone, F.~Gonzalez, A.~Hees, C.~Le Poncin-Lafitte and F.~Meynadier, \textit{et al.}
``Gravitational Redshift Test Using Eccentric Galileo Satellites,''
\journal{10.1103/PhysRevLett.121.231101}{Phys. Rev. Lett. \textbf{121}, 231101 (2018)}
\arxiv{1906.06161}{[gr-qc]}


\bibitem{Herrmann:2018rva}
S.~Herrmann, F.~Finke, M.~L{\"u}lf, O.~Kichakova, D.~Puetzfeld, D.~Knickmann, M.~List, B.~Rievers, G.~Giorgi and C.~G{\"u}nther, \textit{et al.}
``Test of the Gravitational Redshift with Galileo Satellites in an Eccentric Orbit,''
\journal{10.1103/PhysRevLett.121.231102}{Phys. Rev. Lett. \textbf{121}, 231102 (2018)}
\arxiv{1812.09161}{[gr-qc]}





\bibitem{RoviraGarcia:2021DBD}
A.~Rovira-Garcia, J.~M.~Juan, J.~Sanz, G.~Gonzalez-Casado, J.~Ventura-Traveset, L.~Cacciapuoti and E.~Schoenemann,
``Removing day-boundary discontinuities on GNSS clock estimates: methodology and results,''
\journal{10.1007/s10291-021-01085-3}{GPS Solutions \textbf{25}, no.2, 35 (2021)}

\bibitem{Shi:2019ClockDiscontinuity}
C.~Shi, S.~Guo, S.~Gu, X.~Yang, X.~Gong, Z.~Deng, M.~Ge and H.~Schuh,
``Multi-GNSS satellite clock estimation constrained with oscillator noise model in the existence of data discontinuity,''
\journal{10.1007/s00190-018-1178-3}{J. Geod. \textbf{93}, no.4, 515-528 (2019)}

\bibitem{Hauschild:2013GNSSClock}
A.~Hauschild, O.~Montenbruck and P.~Steigenberger,
``Short-term analysis of GNSS clocks,''
\journal{10.1007/s10291-012-0278-4}{GPS Solutions \textbf{17}, no.3, 295-307 (2013)}

\bibitem{Huang:2013GPSClock}
G.~Huang, Q.~Zhang, H.~Li and W.~Fu,
``Quality variation of GPS satellite clocks on-orbit using IGS clock products,''
\journal{10.1016/j.asr.2012.09.041}{Adv. Space Res. \textbf{51}, no.6, 978-987 (2013)}

\bibitem{Formichella:2017AcStarkGPS}
V.~Formichella, J.~Camparo and P.~Tavella,
``Influence of the ac-Stark shift on GPS atomic clock timekeeping,''
\journal{10.1063/1.4975071}{Appl. Phys. Lett. \textbf{110}, no.4, 043506 (2017)}



\bibitem{Riley:2008FrequencyStability}
W.~J.~Riley and D.~A.~Howe,
``Handbook of frequency stability analysis,''
\journal{10.6028/NIST.SP.1065}{NIST Special Publication \textbf{1065} (2008)}



\bibitem{Biscarri:2018}
W.~Biscarri, S.~D.~Zhao and R.~J.~Brunner,
``A simple and fast method for computing the Poisson binomial distribution function,''
\journal{10.1016/j.csda.2018.01.007}{Comput. Statist. Data Anal. \textbf{122}, 92--100 (2018)}


\end{thebibliography}
\end{document}